\documentclass[12pt, draftclsnofoot, journal, onecolumn]{IEEEtran}
\usepackage{blindtext}
\usepackage[T1]{fontenc}
\usepackage{textcomp}
\usepackage{amsmath}
\usepackage{amssymb}
\usepackage{bm}
\usepackage{mdwmath}
\usepackage{cases}
\usepackage{graphicx}
\usepackage{xcolor}
\definecolor{myblue}{rgb}{0,0.4980,1} 
\definecolor{myred}{rgb}{0.8706,0.1608,0.0627} 
\usepackage[caption=false,font=footnotesize,subrefformat=parens]{subfig}
\usepackage{makecell}
\usepackage{multirow}
\usepackage{acronym}
\usepackage[nosort,noadjust]{cite}
\usepackage{url}

\usepackage{hyperref}
\hypersetup{%
	bookmarksopen=true,%
	bookmarksnumbered=true,%
	pdfpagemode={UseOutlines},
	pdfstartview={FitH},%
	colorlinks=true,%
	linkcolor={myred},%
	citecolor={cyan}}
\usepackage[noend]{algpseudocode}
\newcounter{MYalgorithmic}
\renewcommand{\theMYalgorithmic}{\arabic{MYalgorithmic}}
\newcommand{\algcaption}[1]{%
	\refstepcounter{MYalgorithmic}%
	\textbf{Algorithm}~\textbf{\theMYalgorithmic}.~#1}
\newenvironment{MYalgorithmic}[5]
{
	\hrule height 1.2pt
	\vspace{3pt}
	#1{#2}%
	#3{#4}
	\vspace{3pt}
	\hrule height 0.5pt
	\vspace{3pt}
	#5
}
{
	\vspace{3pt}
	\hrule height 0.5pt
}

\newcounter{MYitem}[MYalgorithmic]
\renewcommand{\theMYitem}{\arabic{MYitem}}
\newcommand{\algitem}{%
	\refstepcounter{MYitem}%
	\textbf{\theMYitem)}}

\makeatletter
\newcommand{\MYlabel}[1]{\def\@currentlabel{\theALG@line}\label{#1}}
\makeatother

\newtheorem{problem}{\textbf{Problem}}

\newtheorem{theorem}{\textbf{Theorem}}

\newcommand{\upperroman}[1]{\uppercase\expandafter{\romannumeral#1}}

\newcommand{\myvec}[1]{\bm{\mathrm{#1}}}
\newcommand{\myunit}[1]{$ \mathrm{#1} $}
\newcommand{\myexp}{\mathrm{e}}
\DeclareMathOperator{\sbjto}{s.t.}

\newcommand{\MYnewpage}{%
	\ifCLASSOPTIONonecolumn%
		\ifCLASSOPTIONjournal%
			\typeout{The onecolumn journal mode.}%
			\newpage%
		\fi%
	\fi}
\begin{document}
\ifCLASSOPTIONonecolumn
	\typeout{The onecolumn mode.}
	\title{\LARGE Joint Frame Design and Resource Allocation for Ultra-Reliable and Low-Latency Vehicular Networks}
	\author{Haojun~Yang,~\IEEEmembership{Student Member,~IEEE}, Kuan~Zhang,~\IEEEmembership{Member,~IEEE}, Kan~Zheng,~\IEEEmembership{Senior Member,~IEEE}, and~Yi~Qian,~\IEEEmembership{Fellow,~IEEE}
		\thanks{Haojun~Yang and Kan~Zheng are with the Intelligent Computing and Communications ($ \text{IC}^\text{2} $) Lab, Wireless Signal Processing and Networks (WSPN) Lab, Key Laboratory of Universal Wireless Communications, Ministry of Education, Beijing University of Posts and Telecommunications (BUPT), Beijing, 100876, China (E-mail: \textsf{yanghaojun.yhj@bupt.edu.cn; zkan@bupt.edu.cn}).}
		\thanks{Haojun Yang, Kuan~Zhang and Yi~Qian are with the Department of Electrical and Computer Engineering, University of Nebraska-Lincoln, Omaha, NE 68182, USA  (E-mail: \textsf{haojun.yang@unl.edu; kuan.zhang@unl.edu; yi.qian@unl.edu}).}
	}
\else
	\typeout{The twocolumn mode.}
	\title{Joint Frame Design and Resource Allocation for Ultra-Reliable and Low-Latency Vehicular Networks}
	\author{Haojun~Yang,~\IEEEmembership{Student Member,~IEEE}, Kuan~Zhang,~\IEEEmembership{Member,~IEEE}, Kan~Zheng,~\IEEEmembership{Senior Member,~IEEE}, and~Yi~Qian,~\IEEEmembership{Fellow,~IEEE}
		\thanks{Haojun~Yang and Kan~Zheng are with the Intelligent Computing and Communications ($ \text{IC}^\text{2} $) Lab, Wireless Signal Processing and Networks (WSPN) Lab, Key Laboratory of Universal Wireless Communications, Ministry of Education, Beijing University of Posts and Telecommunications (BUPT), Beijing, 100876, China (E-mail: \textsf{yanghaojun.yhj@bupt.edu.cn; zkan@bupt.edu.cn}).}
		\thanks{Haojun Yang, Kuan~Zhang and Yi~Qian are with the Department of Electrical and Computer Engineering, University of Nebraska-Lincoln, Omaha, NE 68182, USA  (E-mail: \textsf{haojun.yang@unl.edu; kuan.zhang@unl.edu; yi.qian@unl.edu}).}
	}
\fi

\ifCLASSOPTIONonecolumn
	\typeout{The onecolumn mode.}
\else
	\typeout{The twocolumn mode.}
	\markboth{IEEE Transactions on Wireless Communications}{Yang \MakeLowercase{\textit{et al.}}: Joint Frame Design and Resource Allocation for Ultra-Reliable and Low-Latency Vehicular Networks}
\fi

\maketitle

\ifCLASSOPTIONonecolumn
	\typeout{The onecolumn mode.}
	\vspace*{-50pt}
\else
	\typeout{The twocolumn mode.}
\fi
\begin{abstract}
The rapid development of the fifth generation mobile communication systems accelerates the implementation of vehicle-to-everything communications. Compared with the other types of vehicular communications, vehicle-to-vehicle (V2V) communications mainly focus on the exchange of driving safety information with neighboring vehicles, which requires ultra-reliable and low-latency communications (URLLCs). However, the frame size is significantly shortened in V2V URLLCs because of the rigorous latency requirements, and thus the overhead is no longer negligible compared with the payload information from the perspective of size. In this paper, we investigate the frame design and resource allocation for an urban V2V URLLC system in which the uplink cellular resources are reused at the underlay mode. Specifically, we first analyze the lower bounds of performance for V2V pairs and cellular users based on the regular pilot scheme and superimposed pilot scheme. Then, we propose a frame design algorithm and a semi-persistent scheduling algorithm to achieve the optimal frame design and resource allocation with the reasonable complexity. Finally, our simulation results show that the proposed frame design and resource allocation scheme can greatly satisfy the URLLC requirements of V2V pairs and guarantee the communication quality of cellular users.
\end{abstract}

\ifCLASSOPTIONonecolumn
	\typeout{The onecolumn mode.}
	\vspace*{-10pt}
\else
	\typeout{The twocolumn mode.}
\fi
\begin{IEEEkeywords}
Vehicular networks (VNET), vehicle-to-vehicle (V2V), ultra-reliable and low-latency communications (URLLC), finite blocklength theory, massive MIMO.
\end{IEEEkeywords}

\IEEEpeerreviewmaketitle

\MYnewpage


\section{Introduction}
\label{sec:Introduction}

\IEEEPARstart{W}{ith} the remarkable advancements in the fifth generation (5G) mobile communication systems, ultra-reliable and low-latency communications (URLLCs) become the indispensable components of vehicular networks (VNETs)~\cite{adv2015,Sutton2019,YangMag2019}. The performance of emerging vehicle-to-everything (V2X) communications can be greatly improved, especially the latency and reliability for safety-related applications~\cite{zhengsurvey2015}. Compared with the other types of vehicular communications, vehicle-to-vehicle (V2V) communications mainly focus on the exchange of driving safety information with neighboring vehicles, which requires the rigorous latency and reliability. In order to achieve the goal of URLLCs, 3GPP also declares that a general URLLC requirement for one transmission of a packet is $ 10^{-5} $ for 32~\myunit{bytes} with a user plane latency of 1~\myunit{ms}~\cite{3gpp913}. However, there are few studies involving the latency and reliability of the \textit{physical layer} in VNETs at present. Therefore, it is paramount to develop new URLLC techniques for V2V communications.

In general, dedicated short range communication (DSRC) systems and long term evolution (LTE)-related systems are used to support V2V communications. Based on the IEEE 802.11p protocol, DSRC systems adopt the carrier sense multiple access (CSMA) technique to support V2V communications~\cite{Morgan2010,zhengsurvey2015}. However, CSMA has two limitations for latency-sensitive V2V communications, i.e., the unbounded access latency and channel collisions. Furthermore, because of the poor deployment of roadside infrastructures, DSRC systems are being abandoned gradually. By contrast, with the aid of device-to-device (D2D) techniques, LTE-related systems are regarded as the most promising solution to V2V communications~\cite{SunS2016,Seo2016}. A typical operational scenario of D2D-based V2V communications is to reuse radio resources with cellular users (CUEs), namely the underlay mode. Based on such a scenario, a radio resource allocation scheme is proposed for V2V in~\cite{Sun2016} by jointly considering the transmission latency and reliability. However, it is essential to consider the impacts of mobility models in VNETs. As an improvement to the above work, a novel V2V resource allocation scheme is presented in~\cite{Mei2018} based on microscopic mobility model, where the relationship between the queueing latency and reliability is studied. Moreover, some other works investigate the problem of maximizing the spectral efficiency (SE) in the traditional V2V/D2D underlay mode~\cite{Feng2013,Ren2015}.

However, the existing works have the following limitations for V2V URLLCs, i.e.,
\begin{enumerate}
\item \label{limitation1} All global instantaneous channel state information (CSI) is known at the base station (BS). It is not practical to let the BS know the CSI of V2V/D2D channels~\cite{Liu2018}. Additionally, the instantaneous CSI is easy to be outdated and is hard to be exchanged in vehicular communications. Exchanging such the CSI causes large overhead and reduces communication efficiency.

\item The existing works focus on the scenario of single-antenna, and ignore the benefits of multi-antenna, such as the diversity gain used to ensure high reliability~\cite{Johansson2015}. For the massive MIMO in 5G, exploiting the channel hardening phenomenon is also beneficial to formulate the large-scale fading-based optimization problems~\cite{zhengsurvey2015MIMO}.
\end{enumerate}

The recent work in~\cite{Liu2018} analyzes the performance and proposes a power control algorithm for the V2V MIMO system, and thus these two limitations are addressed to some extent. However, the mobility model as well as the latency and reliability requirements are vital for the V2V system. In addition, due to the rigorous latency requirement, the frame size is much shorter in V2V URLLCs. Hence, another crucial limitation for the existing works is that neither the ergodic capacity nor the outage capacity are suitable to characterize the tradeoff among the SE, transmission latency and reliability~\cite{Giuseppe2016}. Finally, for the short frame in V2V URLLCs, the size of the control information (overhead or metadata) is no longer negligible compared to that of the payload information as shown in Fig.~\ref{fig_Frame}. The existing works rarely consider the impact of the overhead from the aspect of the physical layer, which is crucial for the performance of V2V URLLCs. Therefore, how to properly design the frame size and efficiently allocate the radio resources are still challenging for V2V URLLCs.

To tackle the above challenges, taking the two-dimensional macroscopic traffic model into account, we study the frame design and resource allocation for an urban V2V URLLC system in this paper. In particular, based on the superimposed pilot (SP) scheme and regular pilot (RP) scheme, we first analyze the lower bounds of performance for V2V pairs and CUEs. Then, according to the lower bounds analyzed, the joint frame design and resource allocation are studied to satisfy the URLLC requirements of V2V pairs while guaranteeing the signal-to-interference-plus-noise ratio (SINR) quality of CUEs. Finally, two low-complexity algorithms are proposed to achieve the optimal frame design and resource allocation. The main contributions of this paper are summarized as follows.
\begin{itemize}
\item For V2V pairs, we derive the lower bound of the tradeoff among the SE, transmission latency and reliability based on the finite blocklength theory. For CUEs, we derive the lower bound of SINR. Furthermore, based on the two-dimensional macroscopic traffic model, the worst case of the above lower bounds is averaged over the vehicle density and large-scale fading under the maximum interference scenario.

\item Based on the worst-case lower bounds, the optimal frame design is studied, of which the optimization objective is to reduce the transmission latency as much as possible. Meanwhile, it guarantees the amount of transmission information for each V2V pair and ensures the SINR quality of CUEs. An iterative algorithm is proposed to obtain the optimal frame design. Moreover, the feasible region of latency and bandwidth is investigated for the V2V URLLC system.

\item According to the optimal frame design achieved, the optimal radio resource allocation, namely the joint optimization of pilot and signal power, is then investigated. The optimization objective is to maximize the minimum amount of transmission information among all V2V pairs. To overcome Limitation~\ref{limitation1}, a semi-persistent scheduling algorithm is proposed to reduce the signaling overhead and implementation complexity for the V2V URLLC system.

\item Our simulation results show that the proposed frame design and resource allocation can satisfy the URLLC requirements of V2V pairs, while guaranteeing the SINR quality of CUEs. Furthermore, in terms of the latency and SE, the SP scheme is the preferred choice for the V2V URLLC system.
\end{itemize}

The remainder of this paper is organized as follows. First of all, Section~\ref{sec:Model} describes the V2V URLLC system model, and Section~\ref{sec:Analysis} analyzes the performance of the V2V URLLC system. Then, the joint optimization of frame size and resource allocation is studied in Section~\ref{sec:Optimization}. Finally, Section~\ref{sec:Simulation} illustrates the simulation results, while the conclusions are offered in Section~\ref{sec:Conclusion}.

\textit{Notations:} Uppercase boldface letters and lowercase boldface letters denote matrices and vectors, respectively, while $ \myvec{I}_N $ denotes an $ N \times N $ identity matrix. Furthermore, $ (\cdot)^\text{T} $, $ (\cdot)^\text{*} $ and $ (\cdot)^\text{H} $ represent the transpose, conjugate and conjugate transpose of a matrix/vector, respectively, while $ [ \cdot ]_{l} $ denotes the $ l $-th element of a vector. Finally, $ \mathbb{E}(\cdot) $ represents the mathematical expectation, while $ \mathbb{CN}(\mu,\sigma^2) $ is the complex Gaussian distribution with mean $ \mu $ and real/imaginary component variance $ \sigma^2/2 $.

\section{V2V URLLC System Model}
\label{sec:Model}

\subsection{Scenario Description}

\begin{figure}[!t]
	\centering
	\includegraphics[scale=0.45]{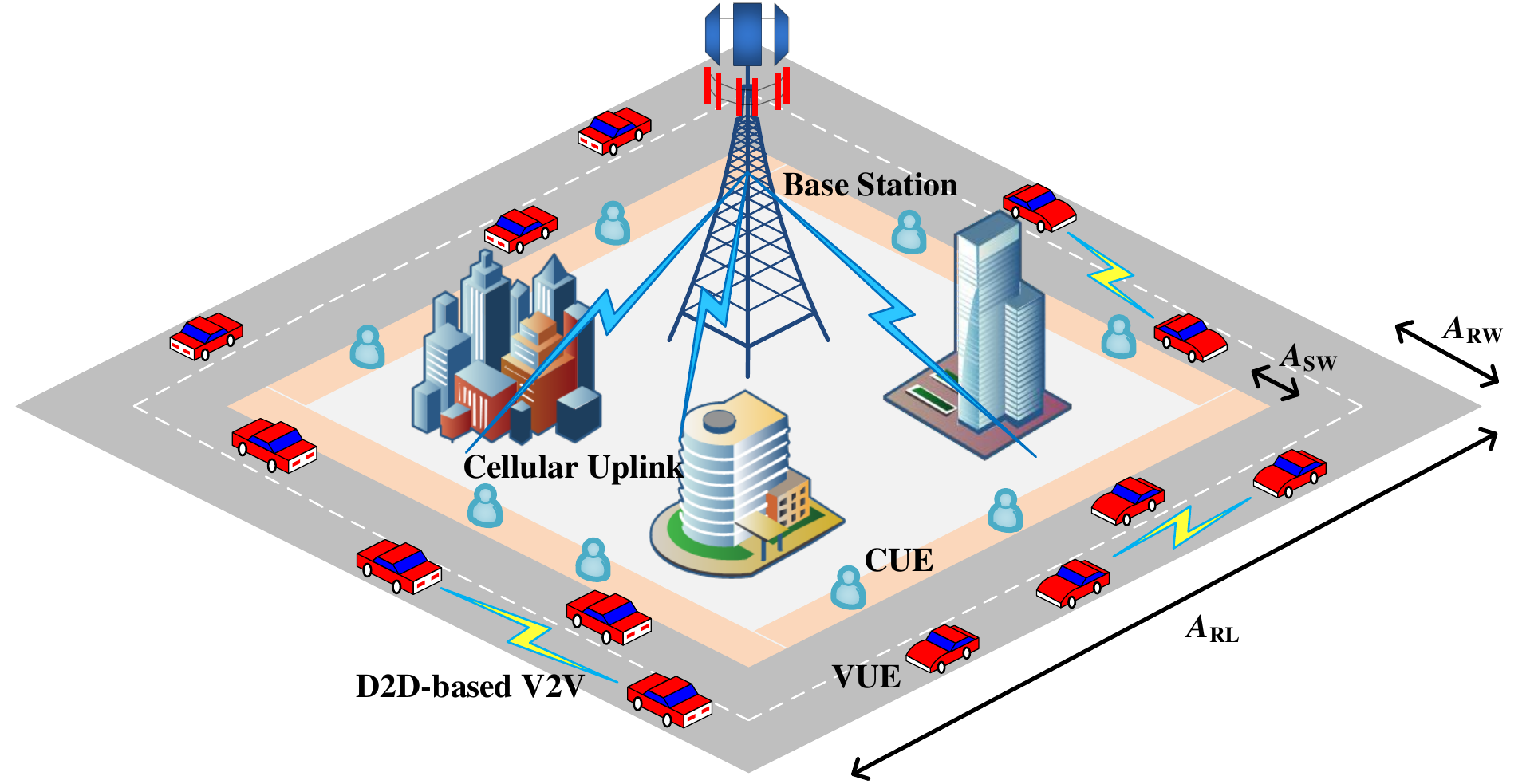}
	\caption{Illustration of system model.}
	\label{fig_Model}
\end{figure}

As shown in Fig.~\ref{fig_Model}, based on the urban grid layout in~\cite[Annex A]{3gpp885}, we consider a single-cell urban V2V URLLC system. Without loss of generality, the length and width of all roads are $ A_\text{RL} $ and $ A_\text{RW} $, respectively. Hence, the length of the square building block is $ A_\text{RL}-A_\text{RW} $. In the building block, $ A_\text{SW} $ is reserved for the sidewalk, where CUEs are uniformly distributed along the sidewalk. In the center of the building block, a BS employs $ M $ antennas and simultaneously communicates with $ K $ single-antenna CUEs. In each road, a total of $ D_u, u \in \{1,2,3,4\} $ V2V pairs reuse the uplink radio resources of CUEs at the underlay mode\footnote{In this paper, we take the underlay mode as a general scenario to analyze and optimize the performance of the V2V URLLC system~\cite{Sun2016,Mei2018,Liu2018}. The overlay mode can be regarded as a special case of the underlay mode. Predictably, if V2V communications operate at the overlay mode, the performance will be improved due to less interferences.}. In each V2V pair, a V2V receiver equipped with $ N $ antennas communicates with a single-antenna V2V transmitter\footnote{As usual, the MISO case is used to simplify the analysis~\cite{Ngo2013,zhengsurvey2015MIMO}. However, it is worth noting that our works can be extended to the MIMO case. Roughly, with respect to the MIMO case, it should first analyze and sum up the SE performance of each stream for each V2V pair, and then the rest works are similar with the MISO case in this paper.}, in order to guarantee the rigorous URLLC requirements. Furthermore, the cooperation of all roads is not considered in this paper, due to the local characteristics of V2V communications. Finally, the system operates in the time-division duplex (TDD) mode.

\subsection{Channel Model}
All channels experience independent flat block-fading, i.e., they remain constant during a coherence block (time-bandwidth product), but change independently from one block to another.
\subsubsection{V2V Channel}
Let $ \myvec{g}^\text{V2V}_{ud,ji} = \sqrt{\beta^\text{V2V}_{ud,ji}} \myvec{h}^\text{V2V}_{ud,ji} \in \mathbb{C}^{N \times 1} $ denote the channel vector spanning from the $ i $-th V2V transmitter on the $ j $-th road to the $ d $-th V2V receiver on the $ u $-th road, where $ \beta^\text{V2V}_{ud,ji} $ and $ \myvec{h}^\text{V2V}_{ud,ji} $ represent the large-scale fading and small-scale fading, respectively. The large-scale fading is given by $ \beta^\text{V2V}_{ud,ji} = \theta (d^\text{V2V}_{ud,ji})^{-\alpha} $, where $ \theta $ is a constant related to the antenna gain and carrier frequency, $ \alpha $ is the path loss exponent, and $ d^\text{V2V}_{ud,ji} $ is the distance between the $ i $-th V2V transmitter on the $ j $-th road and the $ d $-th V2V receiver on the $ u $-th road. Each element of $ \myvec{h}^\text{V2V}_{ud,ji} $ is independent and identically distributed (i.i.d.) complex Gaussian random variable with mean 0 and variance 1, namely $ \myvec{h}^\text{V2V}_{ud,ji} \sim \mathbb{CN}(\myvec{0},\myvec{I}_N) $. Similarly with $ \myvec{g}^\text{V2V}_{ud,ji} $, $ \myvec{g}^\text{V2B}_{ud} = \sqrt{\beta^\text{V2B}_{ud}} \myvec{h}^\text{V2B}_{ud} \in \mathbb{C}^{M \times 1} $ denotes the interference channel vector spanning from the $ d $-th V2V transmitter on the $ u $-th road to the BS.

\subsubsection{CUE Channel}
Similarly with the V2V channel, $ \myvec{g}^\text{C2B}_{k} = \sqrt{\beta^\text{C2B}_{k}} \myvec{h}^\text{C2B}_{k} \in \mathbb{C}^{M \times 1} $ is the channel vector spanning from the $ k $-th CUE to the BS, while $ \myvec{g}^\text{C2V}_{ud,k} = \sqrt{\beta^\text{C2V}_{ud,k}} \myvec{h}^\text{C2V}_{ud,k} \in \mathbb{C}^{N \times 1} $ denotes the interference channel vector spanning from the $ k $-th CUE to the $ d $-th V2V transmitter on the $ u $-th road.

\subsection{Traffic Model}
In general, as the vehicle velocity gradually increases, the fluctuation of small-scale fading becomes more rapid. However, the impact of small-scale fading can be mitigated by exploiting the channel hardening and asymptotic orthogonality of massive MIMO~\cite{Ngo2013,zhengsurvey2015MIMO}. Therefore, the mobility of vehicles (the change of location information) primarily impacts the large-scale fading. It means that the vehicle density (the number of vehicles) and the location information at a certain moment are vital for the performance analysis and optimization of the V2V URLLC system from the aspect of the physical layer. To this end, the first-order macroscopic model is more in line with our needs. In addition, since V2V communications are generally used for traffic safety including lane changing, the first-order \textit{two-dimensional} model considering the mobility in the $ y $-direction (vertical direction) is adopted in this paper. The two-dimensional conservation law is given by $ \partial \rho_u(t,x,y) / \partial t + \partial F_{u,x}(t,x,y) / \partial x + \partial F_{u,y}(t,x,y) / \partial y = 0 $~\cite{Herty2018,trafficflow}, where $ \rho_u $ is the two-dimensional average vehicle density. As the classical first-order model, the Lighthill-Whitham-Richards (LWR) model assumes that the average velocity $ v_u $ only depends on the average vehicle density~\cite{Herty2018,trafficflow}. Then, the average flux (flow rate) of the $ x $-direction and $ y $-direction are given by $ F_{u,x} = \rho_u v_{u,x}(\rho) $ and $ F_{u,y} = \rho_u v_{u,y}(\rho) $. In practice, based on the flow rate observed, traffic control centers can predict and periodically report the density. Therefore, the core component of the macroscopic model is the vehicle density. Moreover, based on the density during a coherent period $ \tilde{\rho}_u (\rho_u = \mathbb{E}[\tilde{\rho}_u]) $, the number of V2V pairs on each road is given by $ D_u = \tilde{\rho}_u A_\text{RL} A_\text{RW} / 2 $ during a period. Finally, the number of V2V pairs on each road follows Poisson distribution~\cite{Zheng2016}, i.e.,
\begin{align}
\label{LWRmodel}
\mathbb{P}\left[ D_u(\rho) = d \right]=\dfrac{\left( \dfrac{\rho_u A_\text{RL}A_\text{RW}}{2} \right)^d}{d!} \myexp^{-\dfrac{\rho_u A_\text{RL}A_\text{RW}}{2}}, \forall u,
\end{align}
where $ \rho_u/2 $ denotes the average V2V pair density.

\section{Performance Analysis of V2V URLLC System}
\label{sec:Analysis}

In this section, we analyze the performance for V2V pairs and CUEs. First of all, two channel estimation schemes are adopted to obtain CSI. Then, for the V2V pairs, the performance of SE, latency and reliability is analyzed based on the finite blocklength theory. Finally, the performance of SINR is analyzed for the CUEs.

\subsection{Channel Estimation of V2V Pairs and CUEs}

\begin{figure}[!t]
	\centering
	\includegraphics[scale=0.45]{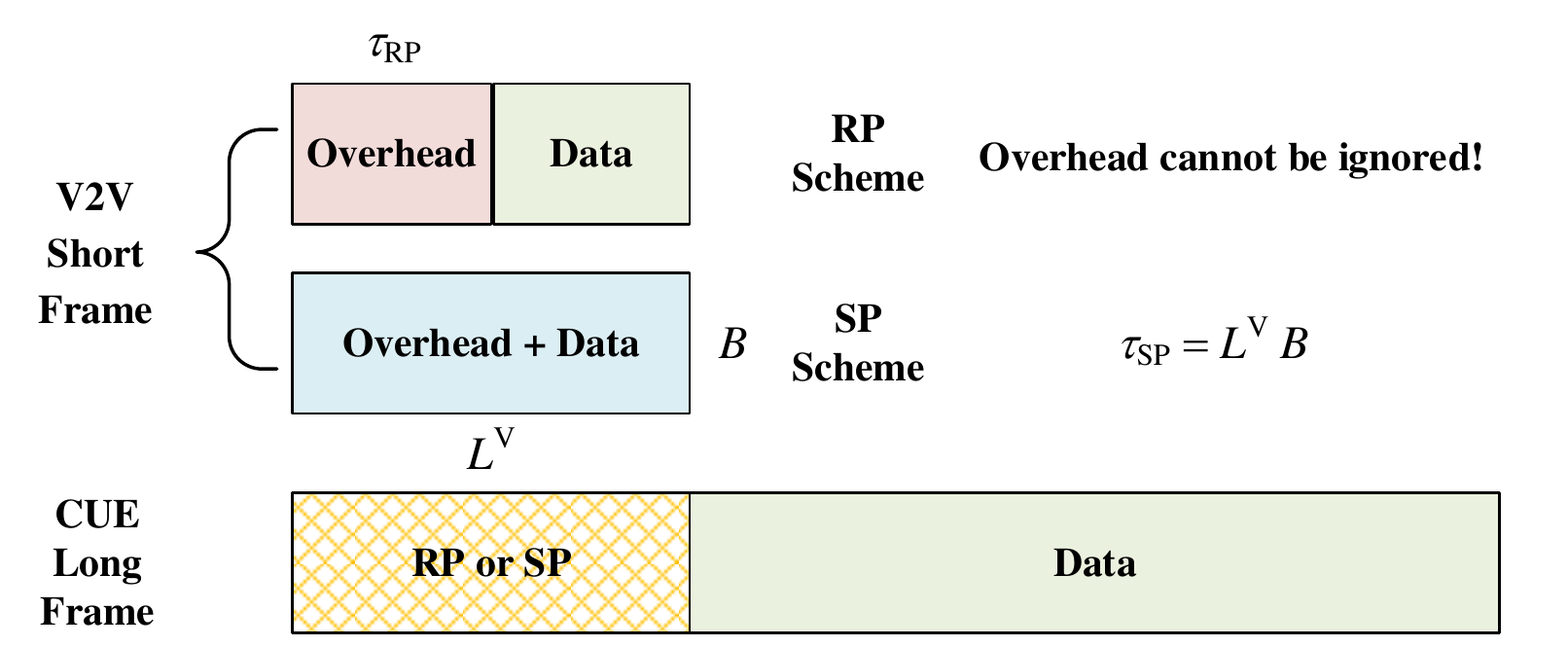}
	\caption{Illustration of two channel estimation schemes.}
	\label{fig_Frame}
\end{figure}

Because vehicles generally have more powerful processing capabilities compared with CUEs, it is reasonable to let V2V communications operate with massive MIMO~\cite{Liu2018,Jiang2018,ChenY2017}, which is beneficial for improving the degrees of freedom to estimate the desired and interference CSI. Furthermore, if the V2V pairs operate with massive MIMO, the effect of semi-persistent scheduling can be achieved, which is detailedly stated in Section~\ref{sec:SPS}. Let $ L^\text{V} $ and $ B $ be the transmission latency and system bandwidth, respectively. Thus $ \tau_\text{SP} = L^\text{V}B $, which is also referred to as the channel uses, represents the number of transmitted symbols or the size of a coherence block.

\subsubsection{Regular Pilot}
Since the RP scheme is often used in the existing wireless systems including the LTE-related systems~\cite{Dahlmanbook}, the RP scheme is considered for comparison in this paper. As shown in Fig.~\ref{fig_Frame}, $ \tau_\text{RP} $ out of $ \tau_\text{SP} $ symbols are utilized for estimating in the RP scheme. For all CUEs, the BS allocates $ K $ different pilots for them from a set of $ \tau_\text{RP} $ ($ K \leqslant \tau_\text{RP} < \tau_\text{SP} $) orthogonal pilots of length $ \tau_\text{RP} $. Let $ \myvec{a}_{k} \in \mathbb{C}^{\tau_\text{RP} \times 1} $ denote the pilot allocated to the $ k $-th CUE, where
\begin{align}
\label{pilotalloc}
\myvec{a}_{k}^\text{H} \myvec{a}_{k^\prime} =
\begin{cases}
\tau_\text{RP} &\text{for}~k^\prime = k,\\
0 &\text{for}~k^\prime \neq k.
\end{cases}
\end{align}
For all V2V transmitters, the pilots are selected randomly and independently from a set of $ \tau_\text{RP} $ orthogonal pilots. In fact, the BS broadcasts the pilot allocation to all V2V pairs as shown in Algorithm~\ref{alg2}. The general case is studied in this paper, namely the random allocation for V2V communications. Let $ \myvec{a}_{ud} \in \mathbb{C}^{\tau_\text{RP} \times 1} $ be the pilot selected by the $ d $-th V2V transmitter on the $ u $-th road. Hence, for arbitrary $ (j,i) \neq (u,d) $ and the $ k $-th CUE, we have 
\begin{align}
\label{dist}
\myvec{a}_{ud}^\text{H} \myvec{a}_{k} = \myvec{a}_{ud}^\text{H}\myvec{a}_{ji} =
\begin{cases}
\tau_\text{RP} &\text{with probability}~\dfrac{1}{\tau_\text{RP}},\\
0 &\text{with probability}~1-\dfrac{1}{\tau_\text{RP}}.
\end{cases}
\end{align}

The received pilot signal $ \myvec{Y}^\text{V,RP}_{ud} \in \mathbb{C}^{N \times \tau_\text{RP}} $ at the $ d $-th V2V receiver on the $ u $-th road is
\begin{align}
\myvec{Y}^\text{V,RP}_{ud} = \sum_{j=1}^{4} \sum_{i=1}^{D_j} \sqrt{q^\text{V}_{ji}}\myvec{g}^\text{V2V}_{ud,ji}\myvec{a}_{ji}^\text{T} + \sum_{k=1}^{K} \sqrt{q^\text{C}_{k}}\myvec{g}^\text{C2V}_{ud,k}\myvec{a}_{k}^\text{T} + \myvec{N}^\text{V,RP}_{ud},
\end{align}
where $ q^\text{V}_{ji} $ and $ q^\text{C}_{k} $ are the pilot power of the $ i $-th V2V transmitter on the $ j $-th road and the $ k $-th CUE, respectively. The i.i.d. additive white Gaussian noise (AWGN) matrix can be written as $ \myvec{N}^\text{V,RP}_{ud} = [\myvec{n}^\text{V,RP}_{ud,1}, \cdots, \myvec{n}^\text{V,RP}_{ud,l}, \cdots, \myvec{n}^\text{V,RP}_{ud,\tau_\text{RP}}] \in \mathbb{C}^{N \times \tau_\text{RP}} $, where $ \myvec{n}^\text{V,RP}_{ud,l} \sim \mathbb{CN}(\myvec{0},\sigma^2\myvec{I}_N), \forall l \in \{1,2,\cdots,\tau_\text{RP}\} $. By multiplying $ \myvec{Y}^\text{V,RP}_{ud} $ with $ \myvec{a}_{ud}^\text{*} / \sqrt{\tau_\text{RP}} $, we have
\begin{align}
\label{RPvec}
\myvec{b}^\text{V2V,RP}_{ud,ud} &= \myvec{Y}^\text{V,RP}_{ud} \dfrac{\myvec{a}_{ud}^\text{*}}{\sqrt{\tau_\text{RP}}} = \sqrt{q^\text{V}_{ud}\tau_\text{RP}}\myvec{g}^\text{V2V}_{ud,ud} + \sum_{i=1, i \neq d}^{D_u} \chi^\text{V,RP}_{ui} \sqrt{q^\text{V}_{ui}\tau_\text{RP}}\myvec{g}^\text{V2V}_{ud,ui} \notag \\
&\quad + \sum_{j=1, j \neq u}^{4} \sum_{i=1}^{D_j} \chi^\text{V,RP}_{ji} \sqrt{q^\text{V}_{ji}\tau_\text{RP}}\myvec{g}^\text{V2V}_{ud,ji} + \sum_{k=1}^{K} \chi^\text{V,RP}_{k} \sqrt{q^\text{C}_{k}\tau_\text{RP}}\myvec{g}^\text{C2V}_{ud,k} + \sum_{l=1}^{\tau_\text{RP}}\myvec{n}^\text{V,RP}_{ud,l}\dfrac{[\myvec{a}_{ud}^\text{*}]_l}{\sqrt{\tau_\text{RP}}},
\end{align}
where binary random variables $ \chi^\text{V,RP}_{ji} \in \{ 0,1 \} $ and $ \chi^\text{V,RP}_{k} \in \{ 0,1 \} $ (the distribution obeys \eqref{dist}) indicate whether the $ i $-th V2V transmitter on the $ j $-th road and the $ k $-th CUE have the same pilot with the $ d $-th V2V transmitter on the $ u $-th road, respectively. With the linear minimum mean squared error (LMMSE) technique~\cite{Kaybook}, the estimate of $ \myvec{g}^\text{V2V}_{ud,ud} $ is given by
\begin{align}
\label{VRPest}
\hat{\myvec{g}}^\text{V2V,RP}_{ud,ud} = \dfrac{\omega^\text{V,RP}_{ud}}{\sqrt{q^\text{V}_{ud}\tau_\text{RP}}} \myvec{b}^\text{V2V,RP}_{ud,ud},
\end{align}
where
\begin{align}
\omega^\text{V,RP}_{ud} = \dfrac{q^\text{V}_{ud}\tau_\text{RP}\beta^\text{V2V}_{ud,ud}}{q^\text{V}_{ud}\tau_\text{RP}\beta^\text{V2V}_{ud,ud}+\sum\limits_{i=1, i \neq d}^{D_u} \chi^\text{RP}_{ui} q^\text{V}_{ui}\tau_\text{RP}\beta^\text{V2V}_{ud,ui} + \sum\limits_{j=1, j \neq u}^{4} \sum\limits_{i=1}^{D_j} \chi^\text{RP}_{ji} q^\text{V}_{ji}\tau_\text{RP}\beta^\text{V2V}_{ud,ji}+ \sum\limits_{k=1}^{K} \chi^\text{RP}_{k} q^\text{C}_{k}\tau_\text{RP}\beta^\text{C2V}_{ud,k} + \sigma^2}.
\end{align}

Analogously, the received pilot signal $ \myvec{Y}^\text{C,RP} \in \mathbb{C}^{M \times \tau_\text{RP}} $ at the BS can be written as
\begin{align}
\myvec{Y}^\text{C,RP} = \sum_{k^\prime=1}^{K} \sqrt{q^\text{C}_{k^\prime}}\myvec{g}^\text{C2B}_{k^\prime}\myvec{a}_{k^\prime}^\text{T} + \sum_{u=1}^{4} \sum_{d=1}^{D_u} \sqrt{q^\text{V}_{ud}}\myvec{g}^\text{V2B}_{ud}\myvec{a}_{ud}^\text{T} + \myvec{N}^\text{C,RP}.
\end{align}
Based on the LMMSE technique, the estimate of $ \myvec{g}^\text{C2B}_{k} $ is given by
\begin{align}
\label{CRPest}
\hat{\myvec{g}}^\text{C2B,RP}_{k} = \dfrac{\omega^\text{C,RP}_{k}}{\sqrt{q^\text{C}_{k}\tau_\text{RP}}} \left[ \sqrt{q^\text{C}_{k}\tau_\text{RP}}\myvec{g}^\text{C2B}_{k} + \sum_{u=1}^{4} \sum_{d=1}^{D_u} \chi^\text{C,RP}_{ud}\sqrt{q^\text{V}_{ud}\tau_\text{RP}}\myvec{g}^\text{V2B}_{ud} + \sum_{l=1}^{\tau_\text{RP}}\myvec{n}^\text{C,RP}_{l}\dfrac{[\myvec{a}_{k}^\text{*}]_l}{\sqrt{\tau_\text{RP}}} \right],
\end{align}
where
\begin{align}
\omega^\text{C,RP}_{k} = \dfrac{q^\text{C}_{k}\tau_\text{RP}\beta^\text{C2B}_{k}}{q^\text{C}_{k}\tau_\text{RP}\beta^\text{C2B}_{k}+\sum\limits_{u=1}^{4} \sum\limits_{d=1}^{D_u} \chi^\text{C,RP}_{ud}q^\text{V}_{ud}\tau_\text{RP}\beta^\text{V2B}_{ud}+\sigma^2}.
\end{align}

\subsubsection{Superimposed Pilot}
Now that the overhead is no longer negligible compared with the payload information from the perspective of size, what can we do to reduce the overhead for the short frame in V2V URLLCs? To this end, the solution proposed by this paper is to send the superposition of pilots and data symbols. Compared to the RP scheme, the SP scheme utilizes $ \tau_\text{SP} $ symbols to send a superposition of pilots and data symbols. Due to the longer pilot length, the sufficient channel samples in the SP scheme are more beneficial to estimate the CSI in the high-speed vehicular environments, which can overcome the negative impacts caused by the Doppler frequency shift. Furthermore, the SP scheme can improve the SE, while reducing the pilot contamination for massive MIMO~\cite{Verenzuela2018,Upadhya2017}. Let $ \myvec{c}_{ud} \in \mathbb{C}^{\tau_\text{SP} \times 1} $ and $ \myvec{c}_{k} \in \mathbb{C}^{\tau_\text{SP} \times 1} $ denote the pilot of the $ d $-th V2V transmitter on the $ u $-th road and the $ k $-th CUE, respectively. Similarly, the pilot allocation rule of the SP scheme follows \eqref{pilotalloc} and \eqref{dist}. Given the data symbol vectors $ \myvec{s}^\text{V}_{ji} \in \mathbb{C}^{\tau_\text{SP} \times 1} $ and $ \myvec{s}^\text{C}_{k} \in \mathbb{C}^{\tau_\text{SP} \times 1} $ with $ \mathbb{E}[ \myvec{s}^\text{V}_{ji} \left( \myvec{s}^\text{V}_{ji} \right)^\text{H} ]=\mathbb{E}[ \myvec{s}^\text{C}_{k} \left( \myvec{s}^\text{C}_{k} \right)^\text{H} ]=\myvec{I}_{\tau_\text{SP}} $, the received pilot signal $ \myvec{Y}^\text{V,SP}_{ud} \in \mathbb{C}^{N \times \tau_\text{SP}} $ at the $ d $-th V2V receiver on the $ u $-th road can be written as
\begin{align}
\label{VSP}
\myvec{Y}^\text{V,SP}_{ud} &= \sum_{j=1}^{4} \sum_{i=1}^{D_j} \sqrt{q^\text{V}_{ji}}\myvec{g}^\text{V2V}_{ud,ji}\myvec{c}_{ji}^\text{T} + \sum_{k=1}^{K} \sqrt{q^\text{C}_{k}}\myvec{g}^\text{C2V}_{ud,k}\myvec{c}_{k}^\text{T} + \sum_{j=1}^{4} \sum_{i=1}^{D_j} \sqrt{p^\text{V}_{ji}}\myvec{g}^\text{V2V}_{ud,ji}\left( \myvec{s}^\text{V}_{ji} \right)^\text{T} \notag \\
&\quad + \sum_{k=1}^{K} \sqrt{p^\text{C}_{k}}\myvec{g}^\text{C2V}_{ud,k}\left( \myvec{s}^\text{C}_{k} \right)^\text{T} + \myvec{N}^\text{V,SP}_{ud},
\end{align}
where $ p^\text{V}_{ji} $ and $ p^\text{C}_{k} $ are the signal power of the $ i $-th V2V transmitter on the $ j $-th road and the $ k $-th CUE, respectively. By multiplying $ \myvec{Y}^\text{V,SP}_{ud} $ with $ \myvec{c}_{ud}^\text{*} / \sqrt{\tau_\text{SP}} $, we have
\begin{align}
\myvec{b}^\text{V2V,SP}_{ud,ud} &= \myvec{Y}^\text{V,SP}_{ud} \dfrac{\myvec{c}_{ud}^\text{*}}{\sqrt{\tau_\text{SP}}} = \sqrt{q^\text{V}_{ud}\tau_\text{SP}}\myvec{g}^\text{V2V}_{ud,ud} + \sum_{i=1, i \neq d}^{D_u} \chi^\text{V,SP}_{ui} \sqrt{q^\text{V}_{ui}\tau_\text{SP}}\myvec{g}^\text{V2V}_{ud,ui} \notag \\
&\quad + \sum_{j=1, j \neq u}^{4} \sum_{i=1}^{D_j} \chi^\text{V,SP}_{ji} \sqrt{q^\text{V}_{ji}\tau_\text{SP}}\myvec{g}^\text{V2V}_{ud,ji} + \sum_{k=1}^{K} \chi^\text{V,SP}_{k} \sqrt{q^\text{C}_{k}\tau_\text{SP}}\myvec{g}^\text{C2V}_{ud,k} + \sum_{j=1}^{4} \sum_{i=1}^{D_j} \sqrt{\dfrac{p^\text{V}_{ji}}{\tau_\text{SP}}}\myvec{g}^\text{V2V}_{ud,ji}\left( \myvec{s}^\text{V}_{ji} \right)^\text{T}\myvec{c}_{ud}^\text{*} \notag \\
&\quad + \sum_{k=1}^{K} \sqrt{\dfrac{p^\text{C}_{k}}{\tau_\text{SP}}}\myvec{g}^\text{C2V}_{ud,k}\left( \myvec{s}^\text{C}_{k} \right)^\text{T}\myvec{c}_{ud}^\text{*} + \sum_{l=1}^{\tau_\text{SP}}\myvec{n}^\text{V,SP}_{ud,l}\dfrac{[\myvec{c}_{ud}^\text{*}]_l}{\sqrt{\tau_\text{SP}}},
\end{align}
where all notations are \textit{similar} with \eqref{RPvec}. With the aid of LMMSE technique, the estimate of $ \myvec{g}^\text{V2V}_{ud,ud} $ is given by
\begin{align}
\label{VSPest}
\hat{\myvec{g}}^\text{V2V,SP}_{ud,ud} = \dfrac{\omega^\text{V,SP}_{ud}}{\sqrt{q^\text{V}_{ud}\tau_\text{SP}}} \myvec{b}^\text{V2V,SP}_{ud,ud} = \dfrac{q^\text{V}_{ud}\tau_\text{SP}\beta^\text{V2V}_{ud,ud}}{\mathsf{DEN}^\text{V,SP}_{ud}\sqrt{q^\text{V}_{ud}\tau_\text{SP}}} \myvec{b}^\text{V2V,SP}_{ud,ud},
\end{align}
where
\begin{align}
\mathsf{DEN}^\text{V,SP}_{ud} &= q^\text{V}_{ud}\tau_\text{SP}\beta^\text{V2V}_{ud,ud}+\sum_{i=1, i \neq d}^{D_u} \chi^\text{SP}_{ui} q^\text{V}_{ui}\tau_\text{SP}\beta^\text{V2V}_{ud,ui} + \sum_{j=1, j \neq u}^{4} \sum_{i=1}^{D_j} \chi^\text{SP}_{ji} q^\text{V}_{ji}\tau_\text{SP}\beta^\text{V2V}_{ud,ji} \notag \\
&\quad + \sum_{k=1}^{K} \chi^\text{SP}_{k} q^\text{C}_{k}\tau_\text{SP}\beta^\text{C2V}_{ud,k} + \sum_{j=1}^{4} \sum_{i=1}^{D_j}p^\text{V}_{ji}\beta^\text{V2V}_{ud,ji} + \sum_{k=1}^{K}p^\text{C}_{k}\beta^\text{C2V}_{ud,k} + \sigma^2.
\end{align}

Likewise, the received pilot signal $ \myvec{Y}^\text{C,SP} \in \mathbb{C}^{M \times \tau_\text{SP}} $ at the BS can be written as
\begin{align}
\label{CSP}
\myvec{Y}^\text{C,SP} &= \sum_{k^\prime=1}^{K} \sqrt{q^\text{C}_{k^\prime}}\myvec{g}^\text{C2B}_{k^\prime}\myvec{c}_{k^\prime}^\text{T} + \sum_{u=1}^{4} \sum_{d=1}^{D_u} \sqrt{q^\text{V}_{ud}}\myvec{g}^\text{V2B}_{ud}\myvec{c}_{ud}^\text{T} + \sum_{k^\prime=1}^{K} \sqrt{p^\text{C}_{k^\prime}}\myvec{g}^\text{C2B}_{k^\prime}\left( \myvec{s}^\text{C}_{k^\prime} \right)^\text{T} \notag \\
&\quad + \sum_{u=1}^{4} \sum_{d=1}^{D_u} \sqrt{p^\text{V}_{ud}}\myvec{g}^\text{V2B}_{ud}\left( \myvec{s}^\text{V}_{ud} \right)^\text{T} + \myvec{N}^\text{C,SP}.
\end{align}
According to the LMMSE technique, the estimate of $ \myvec{g}^\text{C2B}_{k} $ is given by
\begin{align}
\label{CSPest}
\hat{\myvec{g}}^\text{C2B,SP}_{k} &= \dfrac{\omega^\text{C,SP}_{k}}{\sqrt{q^\text{C}_{k}\tau_\text{SP}}} \left[ \sqrt{q^\text{C}_{k}\tau_\text{SP}}\myvec{g}^\text{C2B}_{k} + \sum_{u=1}^{4} \sum_{d=1}^{D_u} \chi^\text{C,SP}_{ud}\sqrt{q^\text{V}_{ud}\tau_\text{SP}}\myvec{g}^\text{V2B}_{ud} + \sum_{k^\prime=1}^{K} \sqrt{\dfrac{p^\text{C}_{k^\prime}}{\tau_\text{SP}}}\myvec{g}^\text{C2B}_{k^\prime}\left( \myvec{s}^\text{C}_{k^\prime} \right)^\text{T}\myvec{c}_{k}^\text{*} \right. \notag \\
&\quad \left. + \sum_{u=1}^{4} \sum_{d=1}^{D_u} \sqrt{\dfrac{p^\text{V}_{ud}}{\tau_\text{SP}}}\myvec{g}^\text{V2B}_{ud}\left( \myvec{s}^\text{V}_{ud} \right)^\text{T}\myvec{c}_{k}^\text{*} + \sum_{l=1}^{\tau_\text{SP}}\myvec{n}^\text{C,SP}_{l}\dfrac{[\myvec{c}_{k}^\text{*}]_l}{\sqrt{\tau_\text{SP}}} \right],
\end{align}
where
\begin{align}
\omega^\text{C,SP}_{k} = \dfrac{q^\text{C}_{k}\tau_\text{SP}\beta^\text{C2B}_{k}}{q^\text{C}_{k}\tau_\text{SP}\beta^\text{C2B}_{k}+\sum\limits_{u=1}^{4} \sum\limits_{d=1}^{D_u} \chi^\text{C,SP}_{ud}q^\text{V}_{ud}\tau_\text{SP}\beta^\text{V2B}_{ud} + \sum\limits_{k^\prime=1}^{K}p^\text{C}_{k^\prime}\beta^\text{C2B}_{k^\prime} + \sum\limits_{u=1}^{4} \sum\limits_{d=1}^{D_u}p^\text{V}_{ud}\beta^\text{V2B}_{ud}+\sigma^2}.
\end{align}

\subsection{Signal Transmission of V2V Pairs and CUEs}

\subsubsection{Regular Pilot}
In the RP scheme, the rest $ \tau_\text{SP} - \tau_\text{RP} $ symbols are used for signal transmission. For the $ d $-th V2V receiver on the $ u $-th road, the received signal $ \myvec{y}^\text{V,RP}_{ud} \in \mathbb{C}^{N \times 1} $ is
\begin{align}
\myvec{y}^\text{V,RP}_{ud} = \sum_{j=1}^{4} \sum_{i=1}^{D_j} \sqrt{p^\text{V}_{ji}}\myvec{g}^\text{V2V}_{ud,ji}s^\text{V}_{ji}+ \sum_{k=1}^{K} \sqrt{p^\text{C}_{k}}\myvec{g}^\text{C2V}_{ud,k}s^\text{C}_{k} + \myvec{z}^\text{V,RP}_{ud},
\end{align}
where the vector $ \myvec{z}^\text{V,SP}_{ud} $ is the i.i.d. AWGN with $ \myvec{z}^\text{V,SP}_{ud} \sim \mathbb{CN}(\myvec{0},\sigma^2\myvec{I}_N) $. $ s^\text{V}_{ji} \in \mathbb{C} $ and $ s^\text{C}_{k} \in \mathbb{C} $ denote the date symbol of the $ i $-th V2V transmitter on the $ j $-th road and the $ k $-th CUE, where $ \mathbb{E}[ s^\text{V}_{ji}\left( s^\text{V}_{ji} \right)^\text{*} ] = \mathbb{E}[ s^\text{C}_{k}\left( s^\text{C}_{k} \right)^\text{*} ] = 1 $. It is widely exploited that low-complexity linear detection techniques are capable of asymptotically attaining optimal performance in massive MIMO. Therefore, we adopt the low-complexity maximum ratio combining (MRC) detection in this paper. Recall from \eqref{VRPest} that the post-processing signal of $ \myvec{y}^\text{V,RP}_{ud} $ can be expressed as $ \hat{s}^\text{V,RP}_{ji} = \left( \hat{\myvec{g}}^\text{V2V,RP}_{ud,ud} \right)^\text{H} \myvec{y}^\text{V,RP}_{ud} $.

Analogously, the received signal $ \myvec{y}^\text{C,RP} \in \mathbb{C}^{M \times 1} $ at the BS is given by
\begin{align}
\myvec{y}^\text{C,RP} = \sum_{k^\prime=1}^{K} \sqrt{p^\text{C}_{k^\prime}}\myvec{g}^\text{C2B}_{k^\prime}s^\text{C}_{k^\prime} + \sum_{u=1}^{4} \sum_{d=1}^{D_u} \sqrt{p^\text{V}_{ud}}\myvec{g}^\text{V2B}_{ud}s^\text{V}_{ud} + \myvec{z}^\text{C,RP}.
\end{align}
With the estimated CSI in \eqref{CRPest}, the post-processing signal of the $ k $-th CUE can be written as $ \hat{s}^\text{C,RP}_{k} = \left( \hat{\myvec{g}}^\text{C2B,RP}_{k} \right)^\text{H} \myvec{y}^\text{C,RP} $.

\subsubsection{Superimposed Pilot}
Since the pilot allocation is known to all V2V receiver and the BS, the received pilots can be perfectly canceled at \eqref{VSP} and \eqref{CSP}, in order to reduce the impact of interferences. For an arbitrary data symbol $ [\myvec{s}^\text{V}_{ud}]_l, l \in \{1,2,\cdots,\tau_\text{SP}\} $ of the $ d $-th V2V transmitter on the $ u $-th road, we have its post-processing signal, i.e.,
\begin{align}
\left[ \hat{\myvec{s}}^\text{V,SP}_{ud} \right]_l &= \left( \hat{\myvec{g}}^\text{V2V,SP}_{ud,ud} \right)^\text{H} \left[ \myvec{Y}^\text{V,SP}_{ud} \right]_l = \sum_{j=1}^{4} \sum_{i=1}^{D_j} \sqrt{p^\text{V}_{ji}}\left[ \myvec{s}^\text{V}_{ji} \right]_l \left( \hat{\myvec{g}}^\text{V2V,SP}_{ud,ud} \right)^\text{H}\myvec{g}^\text{V2V}_{ud,ji} \notag \\
&\quad + \sum_{k=1}^{K} \sqrt{p^\text{C}_{k}}\left[ \myvec{s}^\text{C}_{k} \right]_l \left( \hat{\myvec{g}}^\text{V2V,SP}_{ud,ud} \right)^\text{H}\myvec{g}^\text{C2V}_{ud,k} + \left( \hat{\myvec{g}}^\text{V2V,SP}_{ud,ud} \right)^\text{H}\myvec{n}^\text{V,SP}_{ud,l}.
\end{align}

Similarly, based on the estimated CSI in \eqref{CSPest}, the post-processing signal of the $ k $-th CUE in the SP scheme is given by
\begin{align}
\left[ \hat{\myvec{s}}^\text{C,SP}_{k} \right]_l &= \left( \hat{\myvec{g}}^\text{C2B,SP}_{k} \right)^\text{H} \left[ \myvec{Y}^\text{C,SP} \right]_l = \sum_{k^\prime=1}^{K} \sqrt{p^\text{C}_{k^\prime}}\left[ \myvec{s}^\text{C}_{k^\prime} \right]_l \left( \hat{\myvec{g}}^\text{C2B,SP}_{k} \right)^\text{H}\myvec{g}^\text{C2B}_{k^\prime} \notag \\
&\quad + \sum_{u=1}^{4} \sum_{d=1}^{D_u} \sqrt{p^\text{V}_{ud}}\left[ \myvec{s}^\text{V}_{ud} \right]_l \left( \hat{\myvec{g}}^\text{C2B,SP}_{k} \right)^\text{H}\myvec{g}^\text{V2B}_{ud} + \left( \hat{\myvec{g}}^\text{C2B,SP}_{k} \right)^\text{H}\myvec{n}^\text{C,SP}_{l}.
\end{align}

\subsection{Performance Analysis of V2V Pairs}

\subsubsection{Expression of SINR}
For the RP scheme, recall from \eqref{VRPest} that the estimate of $ \myvec{g}^\text{V2V}_{ud,ud} $ can be rewritten as $ \hat{\myvec{g}}^\text{V2V,RP}_{ud,ud} = \omega^\text{V,RP}_{ud}\myvec{g}^\text{V2V}_{ud,ud} + \omega^\text{V,RP}_{ud}\myvec{e}^\text{V2V,RP}_{ud,ud} $, where
\begin{align}
\myvec{e}^\text{V2V,RP}_{ud,ud} &= \sum_{i=1, i \neq d}^{D_u} \chi^\text{V,RP}_{ui} \sqrt{\dfrac{q^\text{V}_{ui}}{q^\text{V}_{ud}}}\myvec{g}^\text{V2V}_{ud,ui} + \sum_{j=1, j \neq u}^{4} \sum_{i=1}^{D_j} \chi^\text{V,RP}_{ji} \sqrt{\dfrac{q^\text{V}_{ji}}{q^\text{V}_{ud}}}\myvec{g}^\text{V2V}_{ud,ji} + \sum_{k=1}^{K} \chi^\text{V,RP}_{k} \sqrt{\dfrac{q^\text{C}_{k}}{q^\text{V}_{ud}}}\myvec{g}^\text{C2V}_{ud,k} + \dfrac{\sum\limits_{l=1}^{\tau_\text{RP}}\myvec{n}^\text{V,RP}_{ud,l}[\myvec{a}_{ud}^\text{*}]_l}{\tau_\text{RP}\sqrt{q^\text{V}_{ud}}}.
\end{align}
Based on the channel hardening, i.e., $ ( \myvec{g}^\text{V2V}_{ud,ud} )^\text{H}\myvec{g}^\text{V2V}_{ud,ud}/N \rightarrow \beta^\text{V2V}_{ud,ud} $, and the asymptotic channel orthogonality, i.e., $ ( \myvec{g}^\text{V2V}_{ud,ud} )^\text{H}\myvec{g}^\text{V2V}_{ud,ji}/N \rightarrow 0, \forall (j,i) \neq (u,d),~( \myvec{g}^\text{V2V}_{ud,ji} )^\text{H}\myvec{g}^\text{C2V}_{ud,k}/N \rightarrow 0, \forall k, (j,i) $ and $ ( \myvec{g}^\text{V2V}_{ud,ji} )^\text{H}\myvec{n}^\text{V}_{ud,l}/N,~( \myvec{g}^\text{C2V}_{ud,k} )^\text{H}\myvec{n}^\text{V}_{ud,l}/N \rightarrow 0, \forall k, (j,i), l $, for the large number of antennas $ N $, the SINR of the $ d $-th V2V receiver on the $ u $-th road can be approximated by
\begin{align}
\label{VRPgamma}
\gamma^\text{V,RP}_{ud}=\dfrac{p^\text{V}_{ud}\left( \beta^\text{V2V}_{ud,ud} \right)^2}{\sum\limits_{i=1, i \neq d}^{D_u} \chi^\text{V,RP}_{ui}p^\text{V}_{ui} \dfrac{q^\text{V}_{ui}}{q^\text{V}_{ud}}\left( \beta^\text{V2V}_{ud,ui} \right)^2 + \sum\limits_{j=1, j \neq u}^{4} \sum\limits_{i=1}^{D_j} \chi^\text{V,RP}_{ji}p^\text{V}_{ji} \dfrac{q^\text{V}_{ji}}{q^\text{V}_{ud}}\left( \beta^\text{V2V}_{ud,ji} \right)^2 + \sum\limits_{k=1}^{K} \chi^\text{V,RP}_{k}p^\text{C}_{k} \dfrac{q^\text{C}_{k}}{q^\text{V}_{ud}}\left( \beta^\text{C2V}_{ud,k} \right)^2}, N \gg 1,
\end{align}
where the noise term is $ o(1/N) $, hence it can be omitted here.

For the SP scheme, by adopting the similar method, the SINR of the $ d $-th V2V receiver on the $ u $-th road can be approximated by
\begin{align}
\label{VSPgamma}
\gamma^\text{V,SP}_{ud}=\dfrac{p^\text{V}_{ud}\left( \beta^\text{V2V}_{ud,ud} \right)^2}{\mathsf{IF}^\text{V,SP}_{ud}}, N \gg 1,
\end{align}
and
\begin{align}
\mathsf{IF}^\text{V,SP}_{ud} &= \sum_{i=1, i \neq d}^{D_u} \chi^\text{V,SP}_{ui}p^\text{V}_{ui} \dfrac{q^\text{V}_{ui}}{q^\text{V}_{ud}}\left( \beta^\text{V2V}_{ud,ui} \right)^2 + \sum_{j=1, j \neq u}^{4} \sum_{i=1}^{D_j} \chi^\text{V,SP}_{ji}p^\text{V}_{ji} \dfrac{q^\text{V}_{ji}}{q^\text{V}_{ud}}\left( \beta^\text{V2V}_{ud,ji} \right)^2 \notag \\
&\quad + \sum_{k=1}^{K} \chi^\text{V,SP}_{k}p^\text{C}_{k} \dfrac{q^\text{C}_{k}}{q^\text{V}_{ud}}\left( \beta^\text{C2V}_{ud,k} \right)^2+\sum_{j=1}^{4} \sum_{i=1}^{D_j} \dfrac{\left( p^\text{V}_{ji} \right)^2}{\tau_\text{SP}q^\text{V}_{ud}}\left( \beta^\text{V2V}_{ud,ji} \right)^2 + \sum_{k=1}^{K} \dfrac{\left( p^\text{C}_{k} \right)^2}{\tau_\text{SP}q^\text{V}_{ud}}\left( \beta^\text{C2V}_{ud,k} \right)^2.
\end{align}

Comparing \eqref{VSPgamma} with \eqref{VRPgamma}, we find that:
\begin{itemize}
\item For both the RP scheme and SP scheme, the SINR is always equivalent to the signal-to-interference ratio (SIR); and
\item The only difference between $ \gamma^\text{V,RP}_{ud} $ and $ \gamma^\text{V,SP}_{ud} $ is that the denominator of $ \gamma^\text{V,SP}_{ud} $ has more data symbol interferences.
\end{itemize}

\subsubsection{Performance of V2V URLLCs}
To efficiently deal with the latency-sensitive communication problems, the finite blocklength theory is the recent advance on the Shannon coding theorem~\cite{Polyanskiy2010,Hayashi2009}. The classical Shannon formula quantifies the error-free capacity at which information can be transmitted over a band-limited channel in the presence of noise and interferences. However, this capacity can only be approached at the cost of excessive transmission (coding) latency, i.e., $ C(\gamma) = \mathbb{E}\left[ \log_2(1+\gamma) \right] = \lim_{\epsilon \rightarrow 0} C_{\epsilon}(\gamma,\epsilon) = \lim_{\epsilon \rightarrow 0}\lim_{n \rightarrow \infty} R(\gamma,n,\epsilon) $. Therefore, both the ergodic capacity and outage capacity are no longer applicable for the \textit{short frame} in V2V URLLCs. The finite blocklength theory properly approximates $ R(\gamma,n,\epsilon) $, and provides a common expression of $ R(\gamma,n,\epsilon) $~\cite{Polyanskiy2010,Hayashi2009,She2017}, i.e.,
\begin{align}
\label{approx}
R\left( \gamma,L,\epsilon \right) = \mathbb{E} \left[ \log_2\left( 1+\gamma \right)-\sqrt{\dfrac{V}{LB}}Q^{-1}\left( \epsilon \right) \right],
\end{align}
where $ \epsilon $ is the proxy for reliability (generally $ 10^{-5} $ or $ 10^{-6} $ for VNETs), and $Q^{-1}(\cdot) $ denotes the inverse of the Gaussian $ Q $-function. $ L $ is the transmission latency, while $ LB $, which is also referred to as the coding blocklength, represents the number of transmitted symbols. Based on \eqref{approx}, one can conclude that as $ LB $ tends to infinity, \eqref{approx} approaches the ergodic capacity. $ V $ is the so-called channel dispersion. For a complex channel, the channel dispersion is given by
\begin{align}
V = \left( 1- \dfrac{1}{(1+\gamma)^2} \right) \left( \log_2 \myexp \right)^2.
\end{align}
In the high SINR region (greater than 10 dB), the channel dispersion can be approximated by $ V=(\log_2 \myexp)^2 $, while in the low SINR region we have $ 0< V < (\log_2 \myexp)^2 $~\cite{Sebastian2015}. Normally, the requirement that $ \gamma \geqslant $ 10~\myunit{dB} can be easily satisfied in the systems supporting URLLCs~\cite{Sun2019}. Therefore, we utilize the approximation $ V=(\log_2 \myexp)^2 $ to obtain a standard lower bound of SE in the V2V URLLC system. The theorem about the SE of V2V pairs is illustrated as follows.

\begin{theorem}
\label{theo1}
Let $ \lambda $ be equal to $ \tau_\text{SP}-\tau_\text{RP} $ and $ \tau_\text{SP} $ in the RP scheme and SP scheme, respectively. For the SE of V2V pairs in the urban V2V URLLC system, a \textit{lower} bound averaged over the pilot allocation is given by
\begin{align}
\label{VSE}
\tilde{R}^\text{V}_{ud} = \log_2\left( 1+\Gamma^\text{V}_{ud} \right)-\log_2 \myexp\sqrt{\dfrac{1}{\lambda}}Q^{-1}\left( \epsilon^\text{V}_{ud} \right),
\end{align}
where
\begin{align}
\label{VGamma}
\Gamma^\text{V}_{ud} =
\begin{cases}
\dfrac{\tau_\text{RP}p^\text{V}_{ud}q^\text{V}_{ud}\left( \beta^\text{V2V}_{ud,ud} \right)^2}{\Phi^\text{V}_{ud}}, &\text{for RP},\\
\dfrac{\tau_\text{SP}p^\text{V}_{ud}q^\text{V}_{ud}\left( \beta^\text{V2V}_{ud,ud} \right)^2}{\Phi^\text{V}_{ud}}, &\text{for SP},
\end{cases}
\end{align}
and
\begin{align}
\Phi^\text{V}_{ud} =
\begin{cases}
\sum\limits_{i=1, i \neq d}^{D_u}p^\text{V}_{ui}q^\text{V}_{ui}\left( \beta^\text{V2V}_{ud,ui} \right)^2+\sum\limits_{j=1, j \neq u}^{4} \sum\limits_{i=1}^{D_j}p^\text{V}_{ji}q^\text{V}_{ji}\left( \beta^\text{V2V}_{ud,ji} \right)^2 + \sum\limits_{k=1}^{K}p^\text{C}_{k}q^\text{C}_{k}\left( \beta^\text{C2V}_{ud,k} \right)^2, & \text{for RP}, \\
\sum\limits_{i=1, i \neq d}^{D_u}p^\text{V}_{ui}q^\text{V}_{ui}\left( \beta^\text{V2V}_{ud,ui} \right)^2+\sum\limits_{j=1, j \neq u}^{4} \sum\limits_{i=1}^{D_j}p^\text{V}_{ji}q^\text{V}_{ji}\left( \beta^\text{V2V}_{ud,ji} \right)^2 \\
\qquad + \sum\limits_{k=1}^{K}p^\text{C}_{k}q^\text{C}_{k}\left( \beta^\text{C2V}_{ud,k} \right)^2+\sum\limits_{j=1}^{4} \sum\limits_{i=1}^{D_j}\left( p^\text{V}_{ji} \right)^2 \left( \beta^\text{V2V}_{ud,ji} \right)^2 + \sum\limits_{k=1}^{K}\left( p^\text{C}_{k} \right)^2 \left( \beta^\text{C2V}_{ud,k} \right)^2, & \text{for SP}.
\end{cases}
\end{align}
Further, by considering the \textit{worst case} without resource allocation (namely the maximum interference scenario), the above lower bound averaged over the vehicle density and large-scale fading is given by
\begin{align}
\bar{\tilde{R}}^\text{V}_{ud} = \log_2\left( 1+\bar{\Gamma}^\text{V}_{ud} \right)-\log_2 \myexp\sqrt{\dfrac{1}{\lambda}}Q^{-1}\left( \epsilon^\text{V}_{ud} \right),
\end{align}
where
\begin{align}
\label{aveVGamma}
\bar{\Gamma}^\text{V}_{ud} =
\begin{cases}
\dfrac{2\tau_\text{RP}}{\left( \rho_u S_\text{R} - 2 \right) \Omega^\text{V2V}_{\text{N},1}\Omega^\text{V2V}_{\text{P},1} + \sum\limits_{j} \rho_j S_\text{R} \Omega^\text{V2V}_{\text{N},2}\Omega^\text{V2V}_{\text{P},1} + \rho_{j^\prime} S_\text{R} \Omega^\text{V2V}_{\text{N},3}\Omega^\text{V2V}_{\text{P},1} + 2 K \psi^2 \Omega^\text{C2V}_{\text{N}}\Omega^\text{V2V}_{\text{P},1}}, &\text{for RP},\\
\dfrac{\tau_\text{SP}}{1+\left( \rho_u S_\text{R} - 2 \right) \Omega^\text{V2V}_{\text{N},1}\Omega^\text{V2V}_{\text{P},1} + \sum\limits_{j} \rho_j S_\text{R} \Omega^\text{V2V}_{\text{N},2}\Omega^\text{V2V}_{\text{P},1} + \rho_{j^\prime} S_\text{R} \Omega^\text{V2V}_{\text{N},3}\Omega^\text{V2V}_{\text{P},1} + 2 K \psi^2 \Omega^\text{C2V}_{\text{N}}\Omega^\text{V2V}_{\text{P},1}}, &\text{for SP}.
\end{cases}
\end{align}
$ S_\text{R} = A_\text{RL}A_\text{RW} $ is the area of each road. $ \psi = P^\text{C}_\text{max}/P^\text{V}_\text{max} $ is the ratio of the maximum power for V2V pairs and CUEs. $ j \in \left\lbrace j | j \perp u  \right\rbrace $ and $ j^\prime \in \left\lbrace j | j \| u  \right\rbrace $. $ j \perp u $ represents the fact that the $ j $-th road is connected (perpendicular) to the $ u $-th road, while $ j \| u $ represents the fact that the $ j $-th road parallels to the $ u $-th road. The results of $ \Omega $ are given by Appendix~\ref{app:calc}.
\end{theorem}

\begin{IEEEproof}
See Appendix~\ref{app:theo1}.
\end{IEEEproof}

According to \eqref{aveVGamma}, we find that $ \bar{\Gamma}^\text{V,RP}_{ud} \approx \bar{\Gamma}^\text{V,SP}_{ud} $ when $ 2\tau_\text{RP} = \tau_\text{SP} $, which means that about 50\% symbols are used for the pilot transmission. Therefore, for the V2V URLLC system, the SP scheme is more efficient than the RP scheme, and the SINR of the former is larger.

\subsection{Performance Analysis of CUEs}

Compared to the V2V pairs, the CUEs do not need to operate with the short frame. Therefore, only the performance of SINR is analyzed for the CUEs here. For the RP scheme and the SP scheme, based on the large number of antennas $ M $, the SINRs are given by~\cite{Verenzuela2018}:
\begin{align}
\label{CRPgamma}
\gamma^\text{C,RP}_{k}=\dfrac{p^\text{C}_{k}\left(\beta^\text{C2B}_{k} \right)^2}{\sum\limits_{u=1}^{4} \sum\limits_{d=1}^{D_u} \chi^\text{C,RP}_{ud}p^\text{V}_{ud}\dfrac{q^\text{V}_{ud}}{q^\text{C}_{k}}\left( \beta^\text{V2B}_{ud} \right)^2}, M \gg 1,
\end{align}
and
\begin{align}
\label{CSPgamma}
\gamma^\text{C,SP}_{k}=\dfrac{p^\text{C}_{k}\left(\beta^\text{C2B}_{k} \right)^2}{\sum\limits_{u=1}^{4} \sum\limits_{d=1}^{D_u} \chi^\text{C,SP}_{ud}p^\text{V}_{ud}\dfrac{q^\text{V}_{ud}}{q^\text{C}_{k}}\left( \beta^\text{V2B}_{ud} \right)^2 + \sum\limits_{k^\prime=1}^{K} \dfrac{\left( p^\text{C}_{k^\prime} \right)^2}{\tau_\text{SP}q^\text{C}_{k}}\left( \beta^\text{C2B}_{k^\prime} \right)^2 + \sum\limits_{u=1}^{4} \sum\limits_{d=1}^{D_u}\dfrac{\left( p^\text{V}_{ud} \right)^2}{\tau_\text{SP}q^\text{C}_{k}}\left( \beta^\text{V2B}_{ud} \right)^2}, M \gg 1.
\end{align}

According to \eqref{CRPgamma} and \eqref{CSPgamma}, we find that the CUEs have less interferences in the denominators of $ \gamma^\text{C,RP}_{k} $ and $ \gamma^\text{C,SP}_{k} $ than those of the V2V pairs, due to the CUEs' orthogonal pilot allocation.


\begin{theorem}
\label{theo2}
For the urban V2V URLLC system operating with the underlay mode, a \textit{lower} bound of the CUEs' SINR averaged over the pilot allocation is given by~\cite{Verenzuela2018,Feng2018}:
\begin{align}
\label{CGamma}
\Gamma^\text{C}_{k} =
\begin{cases}
\dfrac{\tau_\text{RP}p^\text{C}_{k}q^\text{C}_{k}\left(\beta^\text{C2B}_{k} \right)^2}{\sum\limits_{u=1}^{4} \sum\limits_{d=1}^{D_u} p^\text{V}_{ud}q^\text{V}_{ud}\left( \beta^\text{V2B}_{ud} \right)^2}, &\text{for RP},\\
\dfrac{\tau_\text{SP}p^\text{C}_{k}q^\text{C}_{k}\left(\beta^\text{C2B}_{k} \right)^2}{\sum\limits_{u=1}^{4} \sum\limits_{d=1}^{D_u} p^\text{V}_{ud}q^\text{V}_{ud}\left( \beta^\text{V2B}_{ud} \right)^2 + \sum\limits_{k^\prime=1}^{K}\left( p^\text{C}_{k^\prime} \right)^2 \left( \beta^\text{C2B}_{k^\prime} \right)^2 + \sum\limits_{u=1}^{4} \sum\limits_{d=1}^{D_u}\left( p^\text{V}_{ud} \right)^2\left( \beta^\text{V2B}_{ud} \right)^2}, &\text{for SP}.
\end{cases}
\end{align}
Further, by considering the \textit{worst case} without resource allocation, the above lower bound averaged over the vehicle density and large-scale fading is given by
\begin{align}
\label{aveCGamma}
\bar{\Gamma}^\text{C}_{k} =
\begin{cases}
\dfrac{2\tau_\text{RP}\psi^2}{\sum\limits_{u=1}^{4} \rho_u S_\text{R} \Omega^\text{V2B}_{\text{N}}\Omega^\text{C2B}_{\text{P}}}, &\text{for RP},\\
\dfrac{\tau_\text{SP}\psi^2}{\sum\limits_{u=1}^{4} \rho_u S_\text{R} \Omega^\text{V2B}_{\text{N}}\Omega^\text{C2B}_{\text{P}} +K\psi^2\Omega^\text{C2B}_{\text{N}}\Omega^\text{C2B}_{\text{P}}}, &\text{for SP},
\end{cases}
\end{align}
where all notations are similar with \eqref{aveVGamma}.
\end{theorem}

\begin{IEEEproof}
Since the function $ f(x)=1/x, \forall x>0 $ is convex, Jensen's inequality shows $ \mathbb{E}(1/x) \geqslant 1/\mathbb{E}(x) $. The rest proof is similar with Theorem~\ref{theo1}, and thus is omitted here.
\end{IEEEproof}

Similarly with \eqref{aveVGamma}, \eqref{aveCGamma} also shows that $ \bar{\Gamma}^\text{C,RP}_{K} \leqslant \bar{\Gamma}^\text{C,SP}_{k} $ when $ \tau_\text{RP} / \tau_\text{SP} \leqslant 1/2 $, which means up to 50\% symbols are used for the pilot transmission.

\section{Performance Optimization of V2V URLLC System}
\label{sec:Optimization}

With the aid of Theorem~\ref{theo1} and \ref{theo2}, the performance of the V2V URLLC system is optimized in this section, including the frame size, pilot power and signal power.

\subsection{Optimization of Frame Size}

\subsubsection{Problem Formulation}
For V2V URLLCs, the desired goal is to reduce the transmission latency as much as possible, while guaranteeing the amount of transmission information for each V2V pair and ensuring the SINR quality of CUEs. To this end, the following optimization problem is first formulated based on the worst-case lower bounds.

\begin{problem}[Optimization of Frame Size]
\label{pro1}
Let $ \lambda $ be equal to $ \tau_\text{SP}-\tau_\text{RP} $ and $ \tau_\text{SP} $ in the RP scheme and SP scheme, respectively. Given the average vehicle density $ \rho_u, \forall u $, the road length $ A_\text{RL} $ and the road width $ A_\text{RW} $, the optimization of frame size is formulated as
\begin{subequations}
\label{problem1}
\begin{numcases}{\textbf{\text{P\ref{pro1}:}} \quad}
\label{problem11}
\lambda \bar{\tilde{R}}^\text{V}_{ud}\left( \lambda \right) \geqslant \bar{\Theta}^\text{V}_\text{th}, \forall (u,d),~\text{and} \\
\label{problem12}
\bar{\Gamma}^\text{C}_{k} \geqslant \bar{\Theta}^\text{C}_\text{th}, \forall k,
\end{numcases}
\end{subequations}
where $ \bar{\Theta}^\text{V}_\text{th} $ denotes the threshold of the amount of transmission information for each V2V pair, and $ \bar{\Theta}^\text{C}_\text{th} $ denotes the SINR threshold of each CUE.
\end{problem}

\subsubsection{Solution to Problem~\ref{pro1}}
According to \eqref{aveVGamma} and \eqref{aveCGamma}, we conclude that the SP scheme is always more efficient than the RP scheme when the value of $ \tau_\text{SP} $ is given. Hence, here $ \tau^\prime_\text{SP} $ of the RP scheme is optimized as an independent variable which is different from $ \tau_\text{SP} $ of the SP scheme. The purpose of separate optimization is to let RP gain the same performance with SP in Problem~\ref{pro1}. Let $ \zeta $ represent $ \tau_\text{SP} $ for both RP and SP. Denote $ \eta $ as the ratio of $ \tau_\text{RP} $ in $ \zeta $. Based on Theorem~\ref{theo1}, we have the following general expressions for \eqref{problem11}, i.e.,
\begin{align}
f_\text{RP}\left( \eta,\zeta \right) &= \left( 1-\eta \right) \zeta \log_2\left( 1+\bar{a}^\text{V,RP}_{ud} \eta\zeta \right) - b^\text{V}_{ud}\sqrt{\left( 1-\eta \right) \zeta} \triangleq h_\text{RP}\left( \eta,\zeta \right)-g_\text{RP}\left( \eta,\zeta \right) \geqslant \bar{\Theta}^\text{V}_\text{th}, \\
f_\text{SP}\left( \zeta \right) &= \zeta \log_2\left( 1+\bar{a}^\text{V,SP}_{ud} \zeta \right) - b^\text{V}_{ud}\sqrt{\zeta} \triangleq h_\text{SP}\left( \zeta \right) - g_\text{SP}\left( \zeta \right) \geqslant \bar{\Theta}^\text{V}_\text{th},
\end{align}
where $ \bar{a}^\text{V,RP}_{ud} $ and $ \bar{a}^\text{V,SP}_{ud} $ are the merging coefficients given by Theorem~\ref{theo1}. $ b^\text{V}_{ud} = Q^{-1}( \epsilon^\text{V}_{ud} ) \log_2 \myexp $. First of all, let us analyze the properties of the objective functions. With respect to $ \zeta $, we have 
\begin{align}
\label{propertyzeta}
\begin{cases}
\dfrac{\partial h_\text{RP}\left( \eta,\zeta \right)}{\partial \zeta}, \dfrac{\partial^2 h_\text{RP}\left( \eta,\zeta \right)}{\partial \zeta^2} > 0, \dfrac{\partial g_\text{RP}\left( \eta,\zeta \right)}{\partial \zeta} > 0, \dfrac{\partial^2 g_\text{RP}\left( \eta,\zeta \right)}{\partial \zeta^2} < 0, \\
\dfrac{\partial h_\text{SP}\left( \zeta \right)}{\partial \zeta}, \dfrac{\partial^2 h_\text{SP}\left( \zeta \right)}{\partial \zeta^2} > 0, \dfrac{\partial g_\text{SP}\left( \zeta \right)}{\partial \zeta} > 0, \dfrac{\partial^2 g_\text{SP}\left( \zeta \right)}{\partial \zeta^2} < 0,
\end{cases}
\end{align}
which means that $ f_\text{RP}( \eta,\zeta ) $ and $ f_\text{SP}( \zeta ) $ are positive, differentiable, increasing and convex on $ \zeta $. Hence, given $ \eta $ $ h_\text{RP}( \eta,\zeta ) = g_\text{RP}( \eta,\zeta ) + \bar{\Theta}^\text{V}_\text{th} $ and $ h_\text{SP}( \zeta ) = g_\text{SP}( \zeta ) + \bar{\Theta}^\text{V}_\text{th} $ must have the \textit{unique} non-zero solutions $ \zeta^\text{V,RP}_{u, \eta} $ and $ \zeta^\text{V,SP}_{u} $ within the range $ \zeta \in (\zeta_\text{L}, +\infty) $. With respect to $ \eta $ in the RP scheme, we obtain
\begin{align}
\label{propertyeta}
\begin{cases}
\dfrac{\partial h_\text{RP}\left( \eta,\zeta \right)}{\partial \eta} > 0,~\text{for}~0<\eta<\eta^{h_\text{RP}}_*,~\dfrac{\partial h_\text{RP}\left( \eta,\zeta \right)}{\partial \eta} < 0,~\text{for}~\eta^{h_\text{RP}}_*<\eta<1, \\
\dfrac{\partial g_\text{RP}\left( \eta,\zeta \right)}{\partial \eta}, \dfrac{\partial^2 h_\text{RP}\left( \eta,\zeta \right)}{\partial \eta^2}, \dfrac{\partial^2 g_\text{RP}\left( \eta,\zeta \right)}{\partial \eta^2} < 0,
\end{cases}
\end{align}
where $ \eta^{h_\text{RP}}_* = \arg\max_{\eta} \{ h_\text{RP}( \eta,\zeta ) \} $. According to \eqref{propertyeta}, it is hard to intuitively know the other properties of $ f_\text{RP}( \eta,\zeta ) $ on $ \eta $ except the differentiable property. However, an important fact should be noticed, i.e., for the actual communication systems we must have $ f_\text{RP}( \eta,\zeta ) = h_\text{RP}( \eta,\zeta ) - g_\text{RP}( \eta,\zeta ) > 0, \forall \eta \in (\eta_\text{L}, 1) $ and $ h_\text{RP}( 1,\zeta ) = g_\text{RP}( 1,\zeta ) = 0 $. Therefore, $ f_\text{RP}( \eta,\zeta ) $ has the maximum value on $ \eta $ with the given $ \zeta $, which states that $ \partial h_\text{RP}( \eta,\zeta )/\partial \eta = \partial g_\text{RP}( \eta,\zeta )/\partial \eta $ must have the \textit{unique} non-zero solution $ \eta^\text{V}_{u, \zeta} $ within the range of $ \eta \in (\eta_\text{L}, 1) $. At this point, the solutions to \eqref{problem11} can be written as
\begin{align}
\label{optimalpro1VRP}
\left( \eta, \zeta^\text{RP} \right) &\geqslant \max_{u: \zeta^\text{V,RP}_{u,*}} \left\lbrace \exists ! \left( \eta^\text{V}_{u,*}, \zeta^\text{V,RP}_{u,*} \right): f_\text{RP}\left( \eta, \zeta \right) = \bar{\Theta}^\text{V}_\text{th} \right\rbrace, \\
\zeta^\text{SP} &\geqslant \max_u \left\lbrace \zeta^\text{V,SP}_{u,*} \right\rbrace.
\end{align}
By considering together with \eqref{problem12}, the optimal solutions of Problem~\ref{pro1} are given by
\begin{align}
\label{optimalpro1RP}
\left( \eta_*, \zeta^\text{RP}_* \right) &\geqslant
\begin{cases}
\text{RHS of }\eqref{optimalpro1VRP},~\text{for} ~\eta^\text{V}_{*,\text{L}} \zeta^\text{V,RP}_{*,\text{L}} \geqslant \eta^\text{C}_{*,\text{L}} \zeta^\text{C,RP}_{*,\text{L}}, \\
\left( \eta_{*,\text{L}}, \dfrac{\eta^\text{C}_{*,\text{L}} \zeta^\text{C,RP}_{*,\text{L}}}{\eta^\text{V}_{*,\text{L}}} \right),~\text{otherwise},
\end{cases}
\end{align}
\begin{align}
\label{optimalpro1SP}
\zeta^\text{SP}_* &\geqslant \max \left\lbrace \max_u \left\lbrace \zeta^\text{V,SP}_{u,*} \right\rbrace, \dfrac{\sum\limits_{u=1}^{4} \rho_u S_\text{R} \Omega^\text{V2B}_{\text{N}}\Omega^\text{C2B}_{\text{P}} +K\psi^2\Omega^\text{C2B}_{\text{N}}\Omega^\text{C2B}_{\text{P}}}{\left( \bar{\Theta}^\text{C}_\text{th} \right)^{-1} \psi^2} \right\rbrace,
\end{align}
where $ \eta^\text{V}_{*,\text{L}} \zeta^\text{V,RP}_{*,\text{L}} $ is the product of the lower bound of \eqref{optimalpro1VRP}, and $ \eta^\text{C}_{*,\text{L}} \zeta^\text{C,RP}_{*,\text{L}} = \frac{\sum_{u=1}^{4} \rho_u S_\text{R} \Omega^\text{V2B}_{\text{N}}\Omega^\text{C2B}_{\text{P}}}{\left( \bar{\Theta}^\text{C}_\text{th} \right)^{-1} 2\psi^2} $. $ \eta_{*,\text{L}} = \arg\max_{\eta} \{ f_\text{RP}( \eta, \eta^\text{C}_{*,\text{L}} \zeta^\text{C,RP}_{*,\text{L}}/\eta^\text{V}_{*,\text{L}} ) \} $, it can be solved by the loop of Line~\ref{loopdelta} in Algorithm~\ref{alg1}.

\subsubsection{Optimization Algorithm of Frame Size}
By taking the above solutions into account, based on the binary search and Newton's method, Algorithm~\ref{alg1} is proposed for obtaining the optimal frame design. For the SP scheme, only one loop is included, while for the RP scheme, several nested loops are contained. The convergence of Algorithm~\ref{alg1} can be ensured by binary search and Newton's method. Furthermore, the terminal condition $ | \eta^\text{max}_{u,w} - \eta^\text{min}_{u,w} | \leqslant \mu_\eta $ of the RP scheme is equivalent to the condition that $ \exists ! ( \eta^\text{V}_{u,*}, \zeta^\text{V,RP}_{u,*} ): f_\text{RP}( \eta, \zeta ) = \bar{\Theta}^\text{V}_\text{th} $ in Algorithm~\ref{alg1}.

\begin{figure}[!t]
\small
\begin{MYalgorithmic}
\algcaption{Frame design algorithm}
\label{alg1}
\begin{algorithmic}[5]
	\renewcommand{\algorithmicrequire}{\textbf{Initialization:}}
	\Require
	\State \labelitemi~Iterative index $ t=w=\delta=0 $, and maximum iterative tolerances $ \mu_\zeta, \mu_\eta > 0 $.
	\renewcommand{\algorithmicrequire}{\textbf{Frame design for the RP scheme:}}
	\Require
	\State \algitem~Set $ \eta^\text{min}_{u,0} = 0 $ and $ \eta^\text{max}_{u,0} = 1 $.
	\Loop \quad $ (w, \forall u) $
	\State \algitem~Let $ \eta_{w} = \frac{1}{2}\left( \eta^\text{min}_{u,w}+\eta^\text{max}_{u,w} \right) $, $ \zeta^\text{V,RP}_{u,0} = \frac{\eta^\text{C}_{*,\text{L}} \zeta^\text{C,RP}_{*,\text{L}}}{\eta_{w}} $ and $ t = 0 $.
	\Loop \quad $ (t, \forall u) $
	\State \algitem~Calculate $ \zeta^\text{V,RP}_{u,t+1} = \zeta^\text{V,RP}_{u,t} - \frac{f_\text{RP}\left( \eta_{w},\zeta^\text{V,RP}_{u,t} \right)-\bar{\Theta}^\text{V}_\text{th}}{\partial f_\text{RP}\left( \eta_{w},\zeta^\text{V,RP}_{u,t} \right) / \partial \zeta} $.
	\If{$ \left| \zeta^\text{V,RP}_{u,t+1} - \zeta^\text{V,RP}_{u,t} \right| \leqslant \mu_\zeta $}~Update $ \zeta^\text{V,RP}_{u,*} = \zeta^\text{V,RP}_{u,t+1} $, set $ t=t+1 $ and \textbf{\textit{break}}.
	\Else~Set $ t=t+1 $ and \textbf{\textit{continue}}.
	\EndIf
	\EndLoop
	\State \algitem~Let $ \eta^\prime_{w,0} = \eta_{w} $ and $ \delta=0 $.
	\Loop \quad $ (\delta, \forall u) $ \MYlabel{loopdelta}
	\State \algitem~Calculate $ \eta^\prime_{w,\delta+1} = \eta^\prime_{w,\delta} - \frac{\partial f_\text{RP}\left( \eta^\prime_{w,\delta},\zeta^\text{V,RP}_{u,*} \right) / \partial \eta}{\partial^2 f_\text{RP}\left( \eta^\prime_{w,\delta},\zeta^\text{V,RP}_{u,*} \right) / \partial \eta^2} $.
	\If{$ \left| \eta^\prime_{w,\delta+1} - \eta^\prime_{w,\delta} \right| \leqslant \mu_\eta $}~Update $ \eta^\prime_{w,*} = \eta^\prime_{w,\delta+1} $, set $ \delta=\delta+1 $ and \textbf{\textit{break}}.
	\Else~Set $ \delta=\delta+1 $ and \textbf{\textit{continue}}.
	\EndIf
	\EndLoop
	\If{$ \eta^\prime_{w,*} \leqslant \eta_{w} $}~Let $ \eta^\text{max}_{u,w+1} = \eta_{w} $ and $ \eta^\text{min}_{u,w+1} = \eta^\text{min}_{u,w} $.
	\Else~Let $ \eta^\text{min}_{u,w+1} = \eta_{w} $ and $ \eta^\text{max}_{u,w+1} = \eta^\text{max}_{u,w} $.
	\EndIf
	\If{$ \left| \eta^\text{max}_{u,w} - \eta^\text{min}_{u,w} \right| \leqslant \mu_\eta $}~Update $ \eta^\text{V}_{u,*} = \eta^\prime_{w,*} $, set $ w=w+1 $ and \textbf{\textit{break}}.
	\Else~Set $ w=w+1 $ and \textbf{\textit{continue}}.
	\EndIf
	\EndLoop
	\State \algitem~Based on \eqref{optimalpro1RP}, the optimal frame design can be obtained for the RP scheme.
	\setcounter{MYitem}{0}
	\renewcommand{\algorithmicrequire}{\textbf{Frame design for the SP scheme:}}
	\Require
	\State \algitem~Let $ \zeta^\text{V,SP}_{u,0} = \frac{\sum\limits_{u=1}^{4} \rho_u S_\text{R} \Omega^\text{V2B}_{\text{N}}\Omega^\text{C2B}_{\text{P}} +K\psi^2\Omega^\text{C2B}_{\text{N}}\Omega^\text{C2B}_{\text{P}}}{\left( \bar{\Theta}^\text{C}_\text{th} \right)^{-1} \psi^2} $.
	\Loop \quad $ (t, \forall u) $
	\State \algitem~Calculate $ \zeta^\text{V,SP}_{u,t+1} = \zeta^\text{V,SP}_{u,t} - \frac{f_\text{SP}\left( \zeta^\text{V,SP}_{u,t} \right)-\bar{\Theta}^\text{V}_\text{th}}{\partial f_\text{SP}\left( \zeta^\text{V,SP}_{u,t} \right)/ \partial \zeta} $.
	\If{$ \left| \zeta^\text{V,SP}_{u,t+1} - \zeta^\text{V,SP}_{u,t} \right| \leqslant \mu_\zeta $}~Update $ \zeta^\text{V,SP}_{u,*} = \zeta^\text{V,SP}_{u,t+1} $, set $ t=t+1 $ and \textbf{\textit{break}}.
	\Else~Set $ t=t+1 $ and \textbf{\textit{continue}}.
	\EndIf
	\EndLoop
	\State \algitem~Based on \eqref{optimalpro1SP}, the optimal frame design can be obtained for the SP scheme.
\end{algorithmic}
\end{MYalgorithmic}
\end{figure}

\subsubsection{Low Latency: Exchange of Bandwidth}
Based on the optimal frame design acquired by Algorithm~\ref{alg1}, we next study the transmission latency. Let $ \zeta^\text{RP}_{*,\text{L}} $ and $ \zeta^\text{SP}_{*,\text{L}} $ be the lower bound of \eqref{optimalpro1RP} and \eqref{optimalpro1SP}, respectively. Recalling from $ \zeta = L^\text{V}B $, we find that the relationship between transmission latency and system bandwidth is reciprocal, i.e., their product is equal to a constant. Hence, the system bandwidth can be increased to reduce the transmission latency, which is the so-called exchange of bandwidth in the low-latency region. However, there is a limit to increase the system bandwidth. It is well known that MIMO-related techniques operate in the narrow band. To reduce the difficulty of channel estimation, MIMO does not usually work under the frequency selective fading channels. To this end, the exchange limit is the coherence bandwidth $ B_\text{C} $. In conclusion, the \textbf{\textit{feasible region}} of latency and bandwidth is given by
\begin{align}
\label{Region}
\mathcal{R} &\triangleq \left\lbrace \left( L^\text{V}, B \right) | L^\text{V}B \geqslant \zeta_{*,\text{L}}, B \leqslant B_\text{C} \right\rbrace, \\
\label{minL}
L^\text{V}_\text{min} &\triangleq \dfrac{\zeta_{*,\text{L}}}{B_\text{C}},
\end{align}
where $ L^\text{V}_\text{min} $ is the minimum transmission latency. $ \zeta_{*,\text{L}} $ is equal to $ \zeta^\text{RP}_{*,\text{L}} $ and $ \zeta^\text{SP}_{*,\text{L}} $ for both RP and SP.

\subsection{Joint Optimization of Pilot and Signal Power}

\subsubsection{Problem Formulation}
Based on the optimal frame design achieved, the optimal joint optimization of pilot and signal power is investigated. Since the desire goal of V2V URLLCs is to let each V2V pair transmit information within a very low latency, the reasonable optimization objective is to maximize the minimum amount of transmission information among all V2V pairs, rather than maximizing the sum of the amount of transmission information. Furthermore, in the high-speed vehicular environments, the small-scale channel fading generally changes very fast, leading the instantaneous CSI easily to be outdated. Exchanging the instantaneous CSI is also not practical for V2V communications. Hence, in the massive MIMO V2V URLLC system, it is expected to formulate the large-scale fading-based allocation problems~\cite{Liu2018}. Recall from \eqref{VSE} that the amount of transmission information for each V2V pair can be rewritten as
\begin{align}
\tilde{I}^\text{V,RP}_{ud} &= \left( 1-\eta \right) \zeta \log_2\left( 1+\Gamma^\text{V,RP}_{ud} \right) - Q^{-1}\left( \epsilon^\text{V}_{ud} \right) \log_2 \myexp \sqrt{\left( 1-\eta \right) \zeta}, \\
\tilde{I}^\text{V,SP}_{ud} &= \zeta \log_2\left( 1+\Gamma^\text{V,SP}_{ud} \right) - Q^{-1}\left( \epsilon^\text{V}_{ud} \right) \log_2 \myexp \sqrt{\zeta}.
\end{align}

\begin{problem}[Max-Min Resource Allocation]
\label{pro2}
Given the total power $ P^\text{V}_\text{max}, P^\text{C}_\text{max} $, the number of V2V pairs on each road and the location information of all V2V pairs and CUEs, the optimization objective is to maximize the minimum amount of transmission information among all V2V pairs, guaranteeing the SINR quality of CUEs at the same time, i.e.,
\begin{subequations}
\label{problem2RP}
\begin{align}
\textbf{\text{P\ref{pro2}-RP:}} \quad \max_{\substack{\myvec{p}^\text{V},\myvec{q}^\text{V} \\ \myvec{p}^\text{C},\myvec{q}^\text{C}}}~ & \min_{(u,d)} \  \left\lbrace \tilde{I}^\text{V,RP}_{ud} \right\rbrace \\
\label{problem2con1RP}
\sbjto~ & \Gamma^\text{C,RP}_{k} \geqslant \Theta^\text{C}_\text{th}, \forall k, \\
\label{problem2con2RP}
& 0 \leqslant p^\text{V}_{ud},q^\text{V}_{ud} \leqslant P^\text{V}_\text{max}, \forall (u,d), \\
\label{problem2con3RP}
& 0 \leqslant p^\text{C}_{k},q^\text{C}_{k} \leqslant P^\text{C}_\text{max}, \forall k,
\end{align}
\end{subequations}
and
\begin{subequations}
\label{problem2SP}
\begin{align}
\textbf{\text{P\ref{pro2}-SP:}} \quad \max_{\substack{\myvec{p}^\text{V},\myvec{q}^\text{V} \\ \myvec{p}^\text{C},\myvec{q}^\text{C}}}~ & \min_{(u,d)} \  \left\lbrace \tilde{I}^\text{V,SP}_{ud} \right\rbrace \\
\label{problem2con1SP}
\sbjto~ & \Gamma^\text{C,SP}_{k} \geqslant \Theta^\text{C}_\text{th}, \forall k, \\
\label{problem2con2SP}
& 0 \leqslant p^\text{V}_{ud},q^\text{V}_{ud} \leqslant \dfrac{1}{2} P^\text{V}_\text{max}, \forall (u,d), \\
\label{problem2con3SP}
& 0 \leqslant p^\text{C}_{k},q^\text{C}_{k} \leqslant \dfrac{1}{2} P^\text{C}_\text{max}, \forall k.
\end{align}
\end{subequations}
\end{problem}

\subsubsection{Solution to Problem~\ref{pro2}}
For the max-min problem, the original problem \eqref{problem2RP} and \eqref{problem2SP} can be transformed with the epigraph form, namely,
\begin{subequations}
\label{problem2}
\begin{align}
\textbf{\text{P\ref{pro2}-epi:}} \quad \max_{\substack{\myvec{p}^\text{V},\myvec{q}^\text{V} \\ \myvec{p}^\text{C},\myvec{q}^\text{C},\phi}}~ & \left\lbrace \phi \right\rbrace \\
\sbjto~~ & \begin{cases}
\tilde{I}^\text{V,RP}_{ud} \geqslant \phi, \forall (u,d), \\
\phi \geqslant 0, \\
\eqref{problem2con1RP} - \eqref{problem2con3RP},
\end{cases} \text{for RP},~\begin{cases}
\tilde{I}^\text{V,SP}_{ud} \geqslant \phi, \forall (u,d), \\
\phi \geqslant 0, \\
\eqref{problem2con1SP} - \eqref{problem2con3SP},
\end{cases} \text{for SP}.
\end{align}
\end{subequations}
Obviously, the transformed problem \eqref{problem2} is non-convex. However, the constraints $ \tilde{I}^\text{V}_{ud} \geqslant \phi $ and $ \Gamma^\text{C}_{k} \geqslant \Theta^\text{C}_\text{th} $ can be rewritten as the geometric form $ \phi^\prime (\Gamma^\text{V}_{ud})^{-1} \leqslant 1 $ and $ \Theta^\text{C}_\text{th} (\Gamma^\text{C}_{k})^{-1} \leqslant 1 $, where
\begin{align}
\phi^\prime =
\begin{cases}
2^{\frac{\phi+Q^{-1}\left( \epsilon^\text{V}_{ud} \right) \log_2 \myexp \sqrt{\left( 1-\eta \right) \zeta}}{\left( 1-\eta \right) \zeta}}-1, &\text{for RP}, \\
2^{\frac{\phi+Q^{-1}\left( \epsilon^\text{V}_{ud} \right) \log_2 \myexp \sqrt{\zeta}}{\zeta}}-1, &\text{for SP}.
\end{cases}
\end{align}
Therefore, \eqref{problem2} can be further transformed into a standard geometric programming (GP) problem, i.e.,
\begin{subequations}
\label{problem2GP}
\begin{align}
\textbf{\text{P\ref{pro2}-GP:}} \quad \max_{\substack{\myvec{p}^\text{V},\myvec{q}^\text{V} \\ \myvec{p}^\text{C},\myvec{q}^\text{C},\phi^\prime}}~ & \left\lbrace \phi^\prime \right\rbrace \\
\sbjto~~ & \begin{cases}
\phi^\prime \left( \Gamma^\text{V,RP}_{ud} \right)^{-1} \leqslant 1, \forall (u,d), \\
\Theta^\text{C}_\text{th} \left( \Gamma^\text{C,RP}_{k} \right)^{-1} \leqslant 1, \forall k, \\
\phi^\prime \geqslant 2^{\frac{Q^{-1}\left( \epsilon^\text{V}_{ud} \right) \log_2 \myexp}{\sqrt{\left( 1-\eta \right) \zeta}}}-1, \\
\eqref{problem2con2RP} - \eqref{problem2con3RP},
\end{cases} \text{for RP},~\begin{cases}
\phi^\prime \left( \Gamma^\text{V,SP}_{ud} \right)^{-1} \leqslant 1, \forall (u,d), \\
\Theta^\text{C}_\text{th} \left( \Gamma^\text{C,SP}_{k} \right)^{-1} \leqslant 1, \forall k, \\
\phi^\prime \geqslant 2^{\frac{Q^{-1}\left( \epsilon^\text{V}_{ud} \right) \log_2 \myexp}{\sqrt{\zeta}}}-1, \\
\eqref{problem2con2SP} - \eqref{problem2con3SP},
\end{cases} \text{for SP}.
\end{align}
\end{subequations}
The standard GP problem can be efficiently solved by the interior-point methods at the polynomial time complexity~\cite{Boyd2007tutorial}. In this paper, we adopt the \textsf{CVX} package with the solver \textsf{MOSEK} to obtain the global optimum~\cite{cvx,gb08}.

\subsection{Semi-Persistent Scheduling Algorithm for V2V URLLC System}
\label{sec:SPS}

So far, we have optimized the frame size, pilot power and signal power. Next, we study how to obtain the effect of semi-persistent scheduling. Specifically, in order to solve Problem~\ref{pro1} and \ref{pro2}, the proposed algorithms depend on the vehicle density and the location information. However, a crucial characteristic has to be considered in VNETs, i.e., compared to the time-scale of CSI fluctuation on the millisecond level, that of the traffic (including the density and location information) fluctuation is higher (urban vehicle velocity 60~\myunit{km/h} $ \approx $ 1.67~\myunit{cm/ms}~\cite{3gpp885}). Therefore, if the resource allocation is executed on the higher time-scale, the effect of semi-persistent scheduling can be achieved, which means that the signaling overhead and implementation complexity are reduced~\cite{SunS2016,Zheng2016,two-stage,YangMag2019}. In addition, with the aid of the artificial intelligence (AI) technologies, the prediction of the density and location information is beneficial to establish the proposed algorithms. In practice, based on the flow rate observed, traffic control centers can use roadside monitoring sensors or real-world traffic datasets to predict and periodically report the density and location information~\cite{Lv2015,LiuZ2018}. For example, based on the emerging deep learning methods, the short-term traffic is predicted for a part of road network in Beijing~\cite{Han2019}.

In conclusion, Algorithm~\ref{alg2} is proposed for describing a semi-persistent scheduling procedure. It is noticed that compared to the fluctuation of the location information of all V2V pairs, that of the number of all V2V pairs is more stable. Hence, the update interval of the frame size can be longer than that of the resource allocation. According to the macroscopic traffic model, the update interval of the resource allocation can be calculated as the coherence time, i,e., $ T_\text{C} = \sqrt{\frac{9}{16\pi f_\text{MD}^2}} = \sqrt{\frac{9v_\text{C}^2}{16\pi f_\text{C}^2 v^2(\rho)}} $, where $ f_\text{MD} $ is the maximum Doppler frequency, $ f_\text{C} $ is the carrier frequency, and $ v_\text{C} $ is the speed of light. $ v(\rho) $ is related to $ v_{u,x}(\rho)  $ and $ v_{u,y}(\rho), \forall u $. In Algorithm~\ref{alg2}, $ T_\text{RA} $ denotes the time duration of executing a resource allocation policy.

For the multi-cell handover, both the density and location information can be shared via the wired links between traffic control centers and all base stations. As long as these two types of information are available, the proposed algorithms can be readily extended to the multi-cell scenarios. Furthermore, 3GPP recently attempts to introduce positioning systems to assist in resource allocation~\cite{SunS2016}. Thus, the proposed algorithms also can be compatible with the existing standardizations for V2X communications. Finally, because of the wired links, the monitoring and sharing of the density and location information may not bring the excessive signaling overhead for radio access networks.

\begin{figure}[!t]
\small
\begin{MYalgorithmic}
\algcaption{Semi-persistent scheduling algorithm for V2V URLLC system}
\label{alg2}
\begin{algorithmic}[5]
	\renewcommand{\algorithmicrequire}{\textbf{Initialization:}}
	\Require
	\State \labelitemi~Periodically reported vehicle density.
	\State \labelitemi~Periodically reported location information of all V2V pairs and CUEs.
	\State \labelitemi~The length and width of road $ A_\text{RL} $ and $ A_\text{RW} $.
	\State \labelitemi~The thresholds of frame design $ \bar{\Theta}^\text{V}_\text{th} $ and $ \bar{\Theta}^\text{C}_\text{th} $ for each V2V pair and CUE.
	\State \labelitemi~The threshold of resource allocation $ \Theta^\text{C}_\text{th} $ for each CUE.
	\State \labelitemi~Total power $ P^\text{V}_\text{max} $ and $ P^\text{C}_\text{max} $.
	\renewcommand{\algorithmicrequire}{\textbf{Semi-persistent scheduling algorithm:}}
	\Require
	\State \algitem~The BS executes Algorithm~\ref{alg1} to obtain the optimal frame design, and further to conceive the pilot set and allocation. \label{sps1}
	\State \algitem~The BS broadcasts the pilot allocation, vehicle density and location information to all V2V pairs. If the current time slot is not on the moment of updating the average vehicle density, only the latter two information is broadcast. \label{sps2}
	\State \algitem~The BS and all V2V pairs simultaneously solve the joint resource allocation of pilot and signal based on Problem~\ref{pro2}.
	\If{$ T_\text{RA} \leqslant T_\text{C} $}
	\State Maintain the current resource allocation policy.
	\Else
	\State go to Step~\ref{sps2} and \textbf{\textit{continue}}.
	\EndIf
	\If{the current time slot is on the moment of updating the average vehicle density}
	\State go to Step~\ref{sps1} and \textbf{\textit{continue}}.
	\EndIf
\end{algorithmic}
\end{MYalgorithmic}
\end{figure}

\subsection{Complexity Analysis}
Based on the maximum iterative tolerance $ \mu_\zeta $ of Algorithm~\ref{alg1}, the associated precision is given by $ Z_\zeta = \log_{10}(\mu_\zeta^{-1}) $-digit. Therefore, the complexity order of Algorithm~\ref{alg1} used for the SP scheme is $ \mathcal{O}(\log_{2} Z_\zeta) $~\cite{Borweinbook}. Moreover, the total number of iterations for the binary search is given by $ \log_{2} (\mu_\eta^{-1}) $, thus the complexity order of Algorithm~\ref{alg1} used for the RP scheme is $ \mathcal{O}[\log_{2} (\mu_\eta^{-1}) \log_{2} (Z_\zeta Z_\eta)] $, where $ Z_\eta = \log_{10}(\mu_\eta^{-1}) $-digit. With respect to Algorithm~\ref{alg2}, besides the complexity imposed by Algorithm~\ref{alg1}, the other part of complexity is determined by solving the GP problem. In general, the GP problem can be readily transformed into a convex problem which can be solved by the interior-point methods at the \textit{polynomial time} complexity~\cite{Boyd2007tutorial}. A common complexity order is given by $ \mathcal{O}(m^{3.5}) $~\cite{Rashtchi2016}, where $ m $ is the number of inequality constraints. Hence, the complexity order of Algorithm~\ref{alg2} used for both the RP scheme and the SP scheme is $ \mathcal{O}[(5\sum_{u=1}^{4} D_u + 5K + 1)^{3.5}] $.

\section{Simulation Results}
\label{sec:Simulation}

\subsection{Simulation Setup}

\begin{table}[!t]
	\setlength{\extrarowheight}{1pt}
	\centering
	\caption{Basic Simulation Parameters}
	\begin{tabular}{ l | r }
		\hline
		\textbf{Parameter} & \textbf{Value} \\
		\hline
		\hline
		Constant $ \theta $ & $ 10^{-3} $ \\ 
		\hline
		Path loss exponent $ \alpha $ & 3 \\
		\hline
		Maximum power of V2V pairs $ P^\text{V}_\text{max} $ & 200~\myunit{mW} \\
		\hline
		\makecell*[l]{Ratio of the maximum power \\ for V2V pairs and CUEs $ \psi $} & 1 \\
		\hline
		\makecell*[l]{Threshold of the amount of \\ transmission information for V2V $ \bar{\Theta}^\text{V}_\text{th} $} & 32~\myunit{Bytes}~\cite{3gpp885} \\
		\hline
		SINR thresholds for CUEs $ \bar{\Theta}^\text{C}_\text{th} $ and $ \Theta^\text{C}_\text{th} $ & 5 and 10~\myunit{dB} \\
		\hline
	\end{tabular}
	\label{table_1}
\end{table}

In the simulations, the length of each road is set as 200~\myunit{m}. There are 2 lanes in each road and the lane width is set as 4~\myunit{m}~\cite{3gpp885}. Hence, the width of each road is 8~\myunit{m}. Furthermore, the width of sidewalk is set as 3~\myunit{m}~\cite{3gpp885}. Based on Appendix~\ref{app:calc}, all V2V pairs and CUEs are uniformly distributed on the roads and sidewalk. Since the average length of a vehicle is 4~\myunit{m}, the distance between the transmitter and receiver of each V2V pair is set as 12~\myunit{m} (3 vehicles)~\cite{Ghazanfari2019}. Moreover, the length of the square protection region is set as the average width of a vehicle, namely 2~\myunit{m}. Other simulation parameters are listed in Table~\ref{table_1}.

\subsection{Simulation Results and Analysis}

\subsubsection{Convergence}

\begin{figure}[!t]
	\centering
	\subfloat[Convergence of $ \zeta $.]{\includegraphics[scale=0.25]{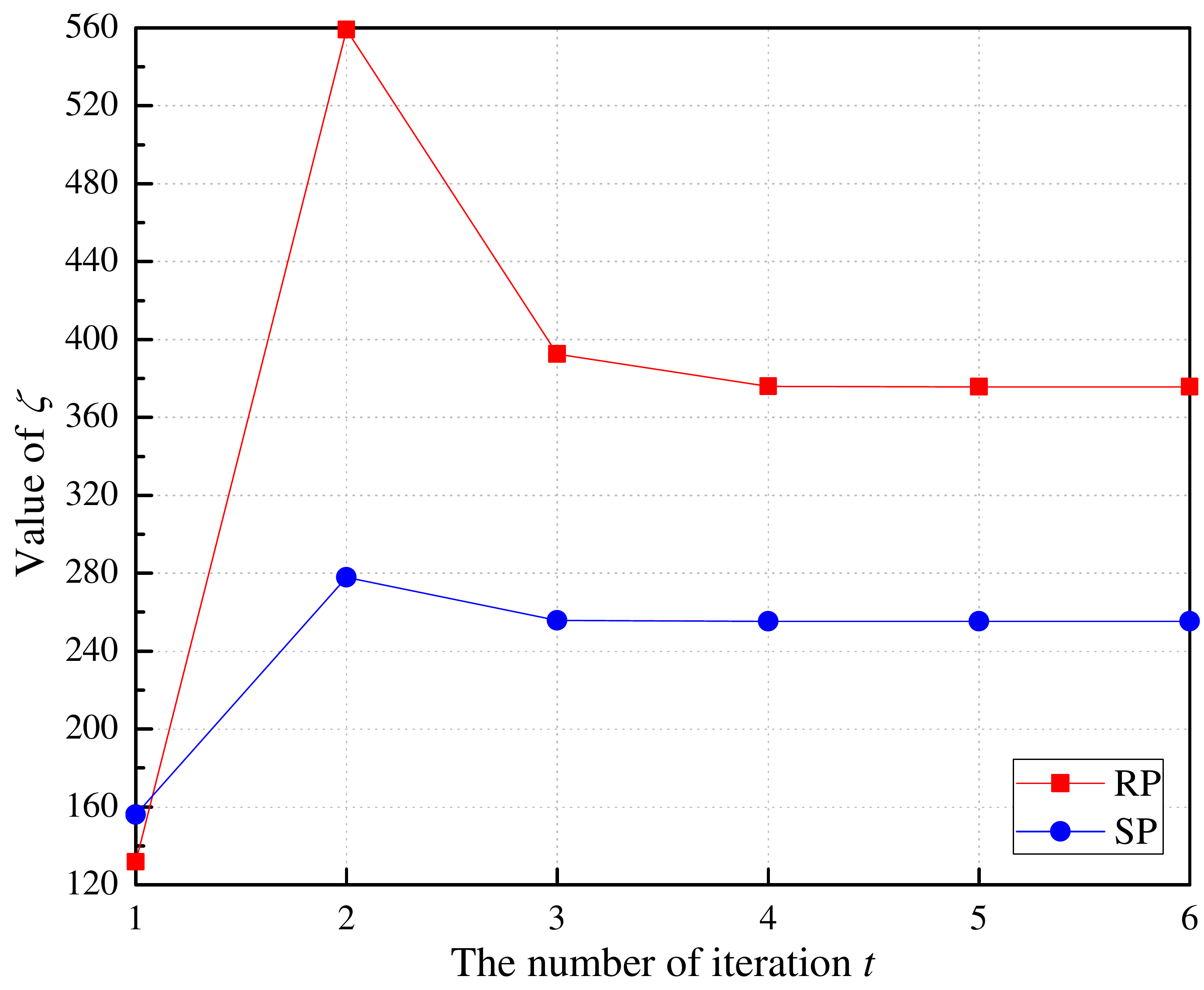}\label{fig_Cov_zeta}}
	\hfil
	\subfloat[Convergence of $ \eta $.]{\includegraphics[scale=0.25]{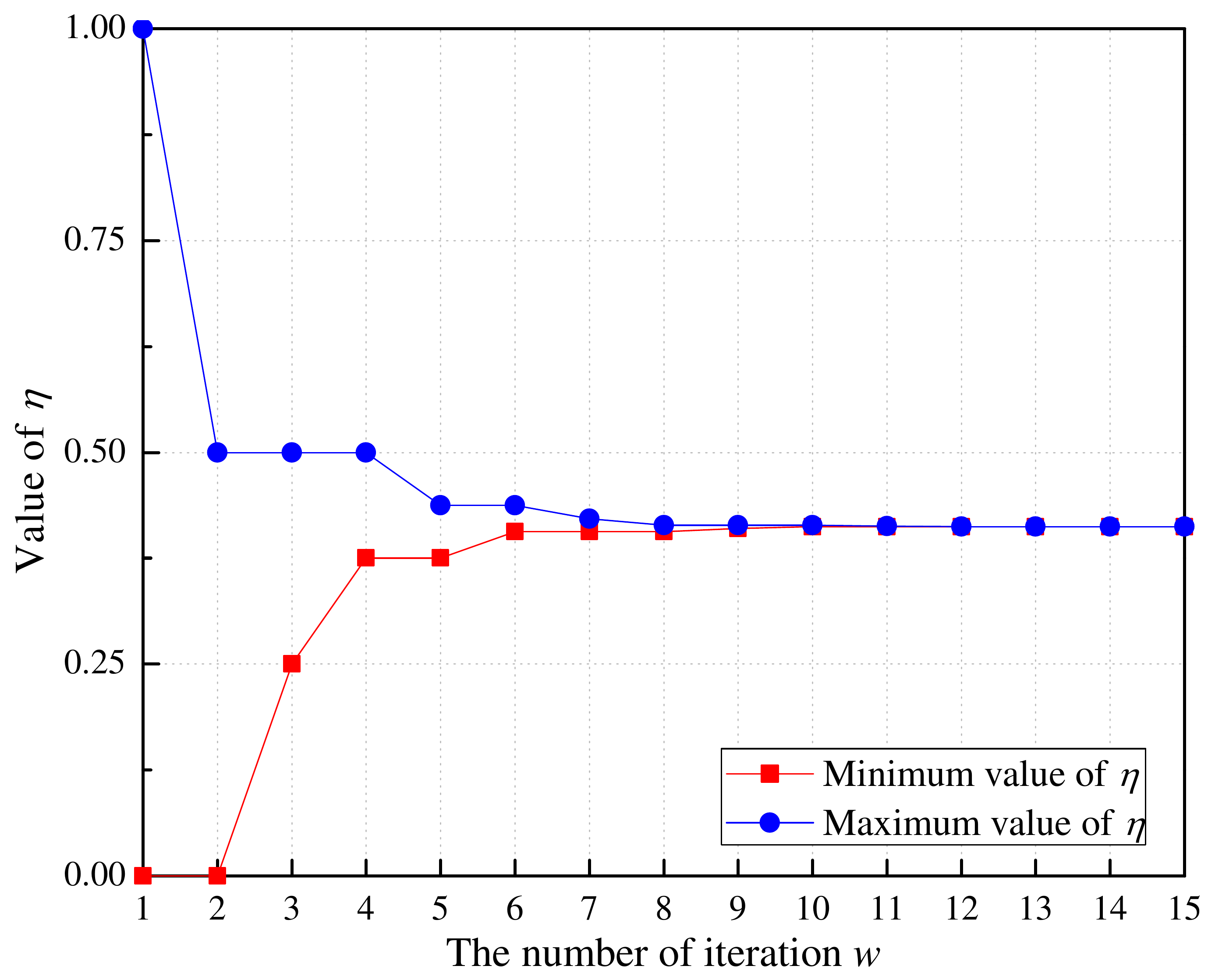}\label{fig_Cov_eta}}
	\caption{Convergence of Algorithm~\ref{alg1}, with $ \mu_\zeta=\mu_\eta = 10^{-4} $, $ \rho_u = 0.005 $ (a total of 16 V2V pairs), $ K = 10 $ and $ \epsilon^\text{V}_{ud} = 10^{-5} $.}
	\label{fig_Cov}
\end{figure}

Fig.~\subref*{fig_Cov_zeta} and \subref*{fig_Cov_eta} illustrate the convergence of $ \zeta $ and $ \eta $, respectively. For both the RP scheme and SP scheme, Fig.~\subref*{fig_Cov_zeta} indicates that as the number of iterations increases, the optimal frame size $ \zeta $ converges to a constant. On the other hand, as shown in Fig.~\subref*{fig_Cov_eta}, as the number of iterations increases, the maximum value of $ \eta $ decreases and the minimum value of $ \eta $ increases. $ \eta^\text{max} $ and $ \eta^\text{min} $ finally converge to one constant, which verifies the convergence of Algorithm~\ref{alg1}. The number of iterations also indicates that Algorithm~\ref{alg1} operates with the reasonable complexity.

\subsubsection{Performance of Optimal Frame Design}

\begin{figure}[!t]
	\centering
	\includegraphics[scale=0.25]{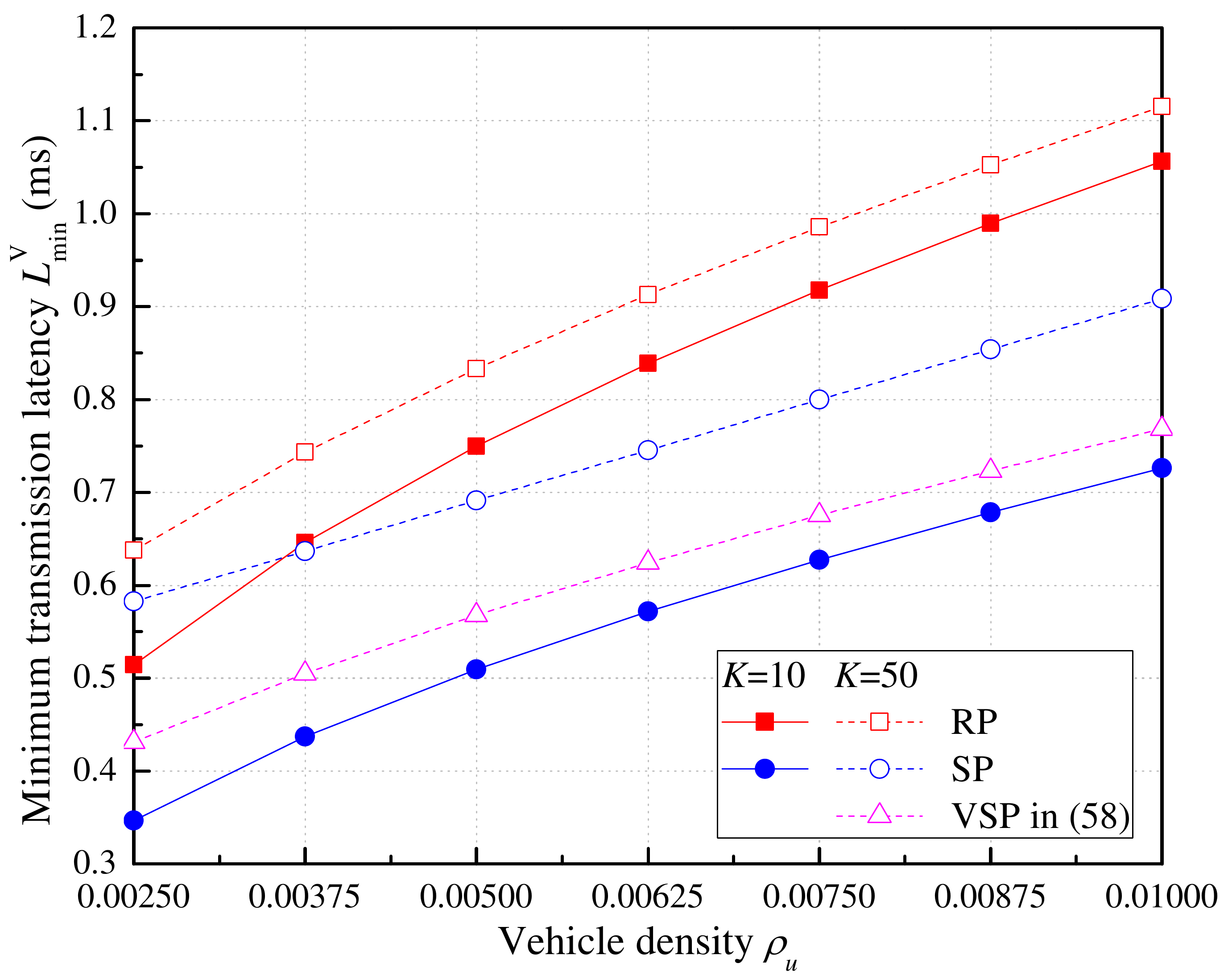}
	\caption{Minimum transmission latency versus vehicle density, with $ \epsilon^\text{V}_{ud} = 10^{-5} $ and $ B_\text{C} = 500 $~\myunit{kHz}.}
	\label{fig_rhovsL}
\end{figure}

Based on the representative value of channel delay spread $ T_\text{D}=1 $~\textmu\myunit{s}, the coherence bandwidth can be calculated as $ B_\text{C}=1/(2T_\text{D})=500 $~\myunit{kHz}~\cite{Tsebook}. According to the coherence bandwidth $ B_\text{C} $ and \eqref{minL}, Fig.~\ref{fig_rhovsL} illustrates the minimum transmission latency versus the vehicle density. For the RP and SP, Fig.~\ref{fig_rhovsL} shows that as the vehicle density increases, the transmission latency increases as well, in order to maintain the performance of the worst case for V2V pairs and CUEs in \eqref{problem1}. As analyzed in \eqref{aveVGamma} and \eqref{aveCGamma}, Fig.~\ref{fig_rhovsL} also shows that the latency performance of SP is always better than that of RP. Furthermore, with the increasing number of CUEs, the latency performance of RP and SP deteriorates. However, the deterioration is acceptable for V2V URLLCs. Finally, the relationship between the minimum transmission latency and vehicle density is linear in the SP scheme when the number of CUEs is equal to 50. This is because \eqref{problem12} dominates the optimal frame size.

\begin{figure}[!t]
	\centering
	\includegraphics[scale=0.25]{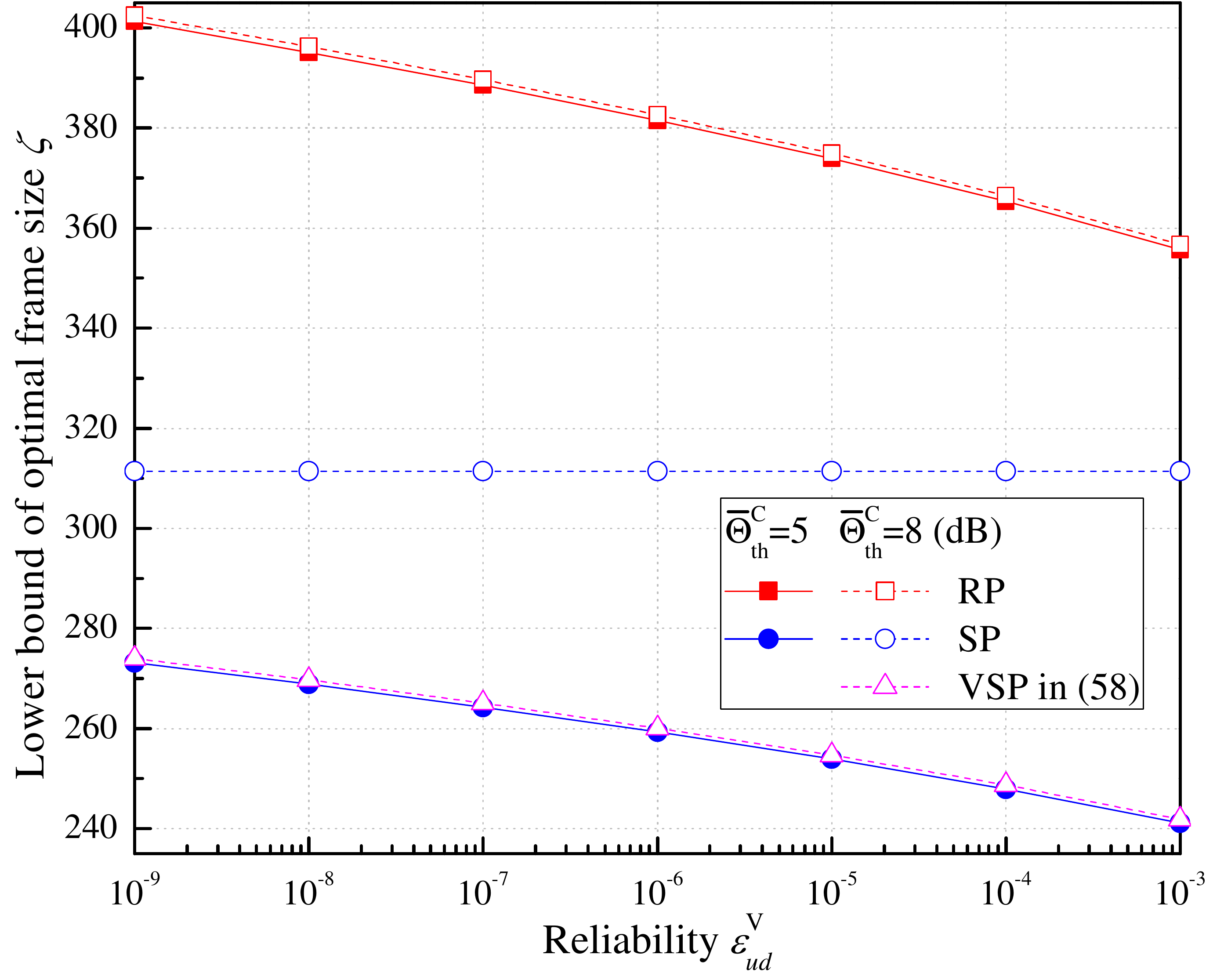}
	\caption{Lower bound of optimal frame size versus reliability, with $ \rho_u = 0.005 $ (a total of 16 V2V pairs) and $ K = 10 $.}
	\label{fig_epsilonvstau}
\end{figure}

Fig.~\ref{fig_epsilonvstau} depicts the lower bound of optimal frame size versus the reliability of V2V URLLCs under two types of CUEs' SINR thresholds. As shown in Fig.~\ref{fig_epsilonvstau}, upon reducing the reliability of V2V URLLCs, the optimal frame size decreases when $ \bar{\Theta}^\text{C}_\text{th} = 5 $~\myunit{dB}, whatever the channel estimation scheme is. However, when the CUEs' SINR threshold is doubled namely $ \bar{\Theta}^\text{C}_\text{th} = 8 $~\myunit{dB}, the optimal frame size of SP maintains a constant, which means that the SP scheme is not sensitive to the reliability of V2V URLLCs. This is because \eqref{problem12} dominates the optimal frame size. By contrast, the improvement of SINR threshold does not have much impact on the RP scheme. Finally, as the reliability of V2V URLLCs increases, the latency of V2V URLLCs is increased by 0.09~\myunit{ms} at most (i.e., $ 45/500 $~\myunit{kHz} in the RP scheme).

\begin{figure}[!t]
	\centering
	\includegraphics[scale=0.35]{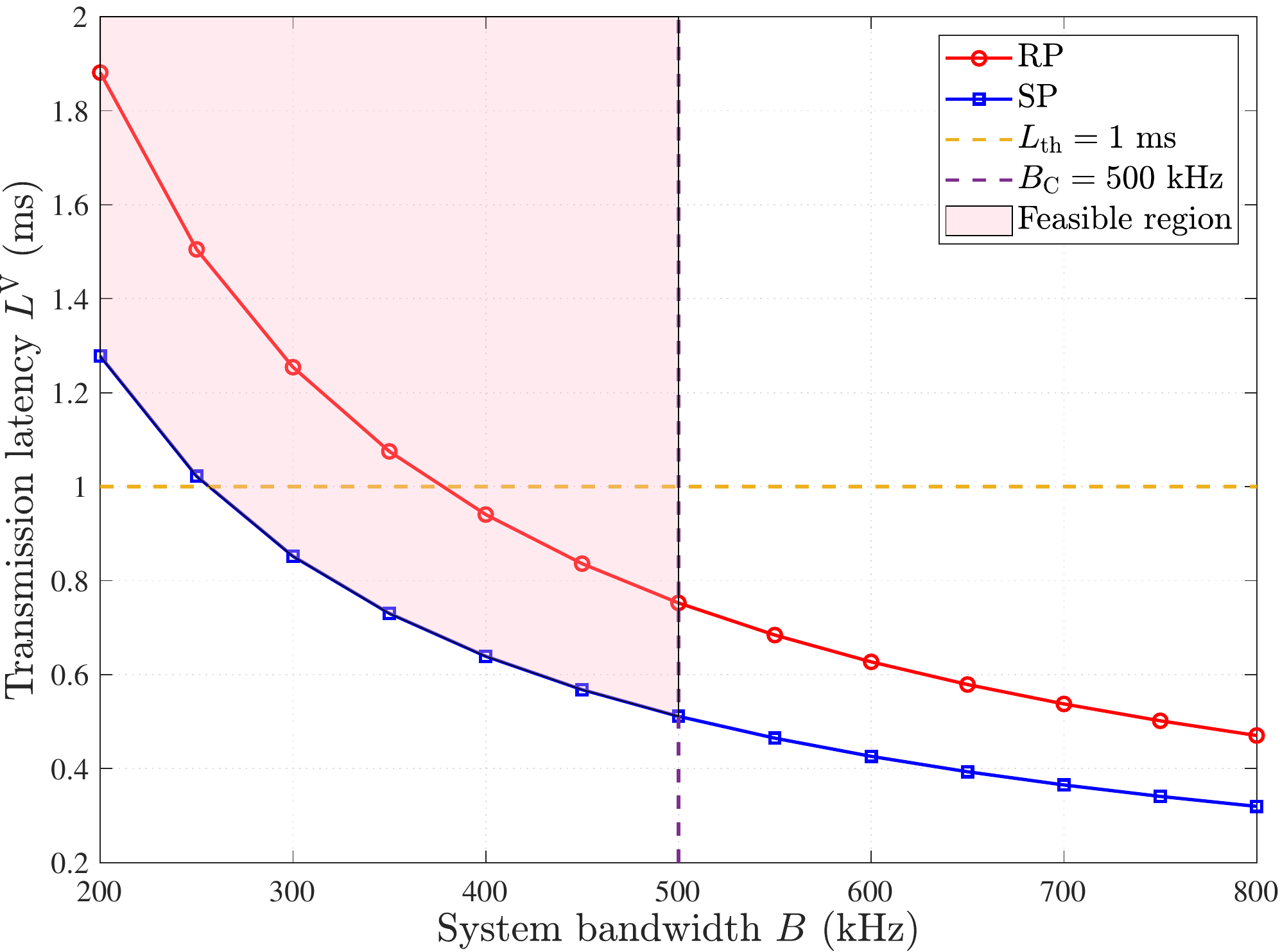}
	\caption{Transmission latency versus system bandwidth, with $ \rho_u = 0.005 $ (a total of 16 V2V pairs), $ K = 10 $ and $ \epsilon^\text{V}_{ud} = 10^{-5} $.}
	\label{fig_BvsL}
\end{figure}

According to \eqref{Region}, Fig.~\ref{fig_BvsL} depicts the transmission latency versus the system bandwidth. As shown in Fig.~\ref{fig_BvsL}, regardless of what the channel estimation scheme is, as the system bandwidth increases, the transmission latency decreases. This is the so-called exchange of bandwidth in the low-latency region. However, there is a limit to increase the system bandwidth, namely the coherence bandwidth. Hence, the exchange of bandwidth only can operate in the feasible region (the pink shaded region). Moreover, if the latency requirement of 3GPP is considered, the feasible region will be reduced. Finally, we also find that when the RP and SP achieve the same transmission performance (i.e., they all satisfy \eqref{problem1}), the latency of SP is lower than that of RP, which means that the SP scheme is a better choice for V2V URLLCs.

\subsubsection{Performance of Optimal Resource Allocation}

\begin{figure}[!t]
	\centering
	\subfloat[CDF of the amount of transmission information for V2V pairs.]{\includegraphics[scale=0.25]{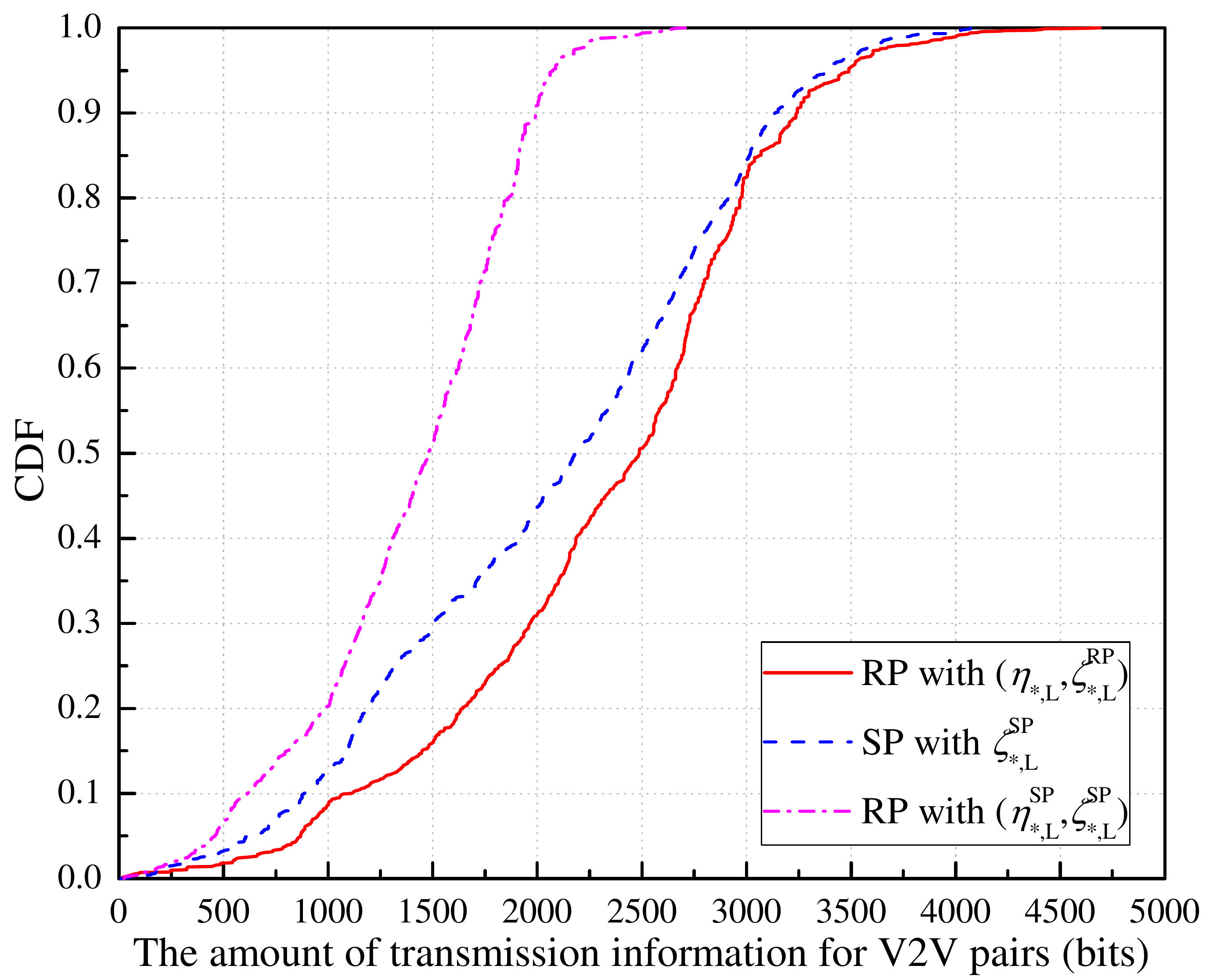}\label{fig_CDFV2V}}
	\hfil
	\subfloat[CDF of uplink SINR for CUEs.]{\includegraphics[scale=0.25]{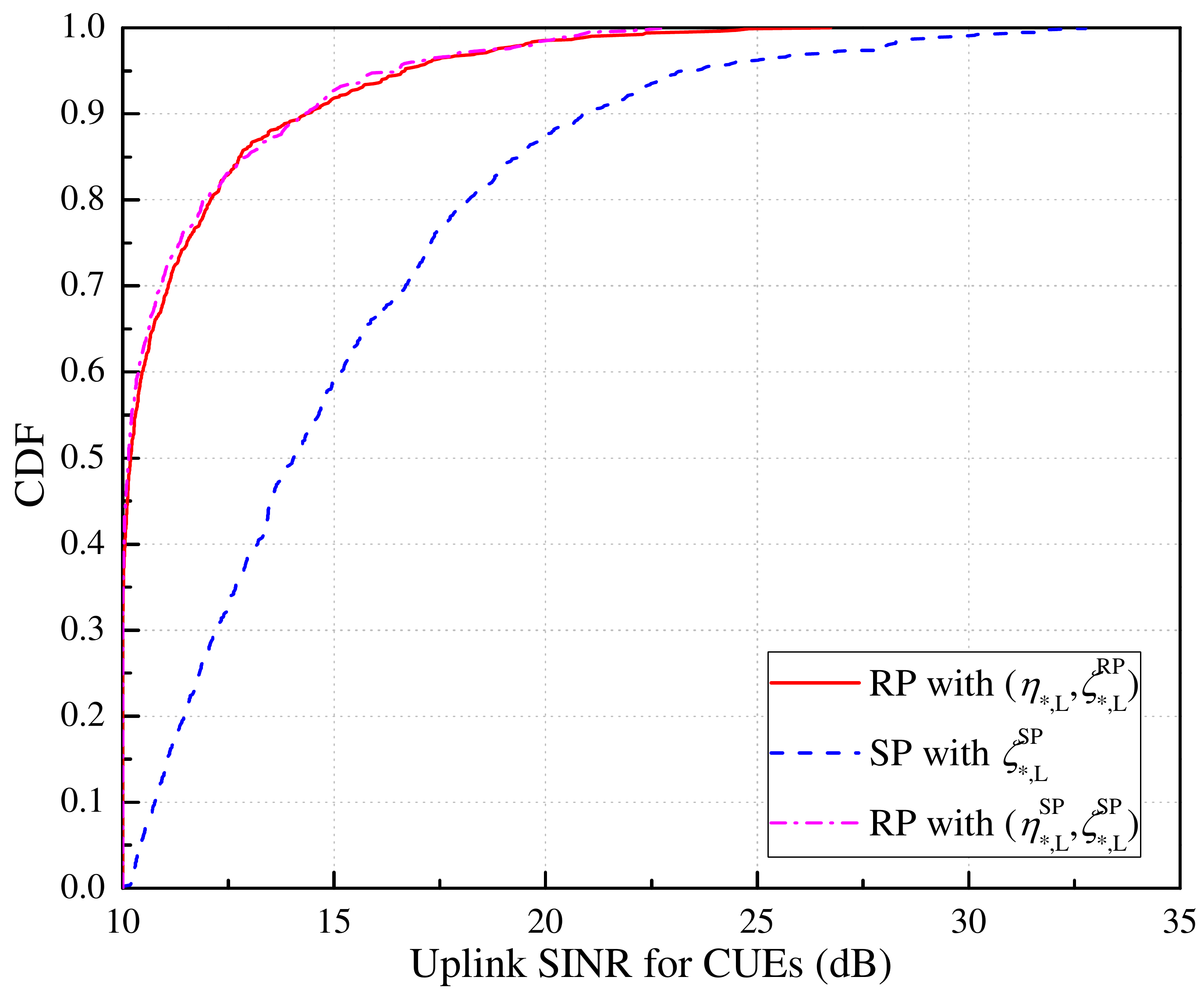}\label{fig_CDFCUE}}
	\caption{CDF performance of V2V URLLC system, with $ \tilde{\rho}_u = 0.0025 $ (a total of 8 V2V pairs), $ K = 4 $ and $ \epsilon^\text{V}_{ud} = 10^{-5} $.}
	\label{fig_CDF}
\end{figure}

Fig.~\subref*{fig_CDFV2V} illustrates the cumulative distribution function (CDF) of the amount of transmission information for V2V pairs. As shown in Fig.~\subref*{fig_CDFV2V}, the RP scheme with $ (\eta_{*,\text{L}}, \zeta^\text{RP}_{*,\text{L}} \approx 246) $ in \eqref{optimalpro1RP} achieves the best performance, while the curve of SP is close to that of RP. This is because the optimal frame size of RP solved by Algorithm~\ref{alg1} is larger than that of SP, which is also evidenced by Fig.~\ref{fig_BvsL}. On the other hand, the SP scheme outperforms the RP scheme with $ (\eta^\text{SP}_{*,\text{L}}, \zeta^\text{SP}_{*,\text{L}} \approx 166) $ in \eqref{optimalpro1SP}, where $ \eta^\text{SP}_{*,\text{L}} $ can be solved by the loop of Line~\ref{loopdelta} in Algorithm~\ref{alg1} based on $ \zeta^\text{SP}_{*,\text{L}} $. In summary, we have the following conclusions: \textit{1)} when the same latency performance is achieved (i.e., when $ \zeta $ is given, and $ 166/500 $~\myunit{kHz} $ = 0.332 $~\myunit{ms}), based on the optimal resource allocation, the amount of transmission information in the SP scheme is larger than that of the RP scheme; \textit{2)} when the same performance of the amount of transmission information is achieved in the worst case (Problem \ref{pro1}), the latency of SP is lower than that of RP ($ 246/500 $~\myunit{kHz} $ = 0.492 $~\myunit{ms}).

Fig.~\subref*{fig_CDFCUE} presents the CDF of uplink SINR for CUEs. As shown in Fig.~\subref*{fig_CDFCUE}, with the SINR threshold $ \Theta^\text{C}_\text{th} = 10 $~\myunit{dB}, the SINR of CUEs are lager than 10~\myunit{dB} under three types of frame size. For the RP scheme, regardless of what the frame size is, two CDF curves of SINR performance are very close. The difference of $ \eta_{*,\text{L}} \zeta^\text{RP}_{*,\text{L}} $ and $ \eta^\text{SP}_{*,\text{L}} \zeta^\text{SP}_{*,\text{L}} $, namely $ \eta_{*,\text{L}} \zeta^\text{RP}_{*,\text{L}} - \eta^\text{SP}_{*,\text{L}} \zeta^\text{SP}_{*,\text{L}} $, is approximately equal to $ 24 $, which means that the slight increase in pilot size has little effect on the SINR of CUEs in the RP scheme. For the SP scheme, the SINR performance improvement is very obvious. This is because the large value of $ \zeta^\text{SP}_{*,\text{L}} \approx 166 $ is conducive to improving $ \Gamma^\text{C,SP}_{k} $ in \eqref{CGamma}. Considering Fig.~\subref*{fig_CDFV2V} and Fig.~\subref*{fig_CDFCUE} together, we find that although the performance of the amount of transmission information for V2V pairs is not optimal in the SP scheme, the latency of V2V pairs and the SINR of CUEs outperform those of RP. Therefore, the SP scheme is more suitable for the urban V2V URLLC system.

\section{Conclusions}
\label{sec:Conclusion}

In order to reduce the overhead for the short frame, this paper studied the problem of jointly optimizing frame size and resource allocation in the urban V2V URLLC system. Utilizing the RP scheme and SP scheme, we first analyzed the lower bounds of performance for V2V pairs and CUEs based on the finite blocklength theory and macroscopic traffic model. Then, with the aid of the lower bounds analyzed, a frame design algorithm and a semi-persistent scheduling algorithm were proposed to achieve optimal frame design and resource allocation. Finally, our simulation results showed that the proposed frame design and resource allocation scheme can greatly satisfy the URLLC requirements of V2V pairs while guaranteeing the SINR quality of CUEs, and stated that the SP scheme was better for the V2V URLLC system.

\appendices

\section{Proof of Theorem~\ref{theo1}}
\label{app:theo1}

With the high SINR and Jensen's inequality ($ f(x)=\log_2(1+1/x), \forall x>0 $ is convex), recall from \eqref{approx} that the SE lower bound of the $ d $-th V2V receiver on the $ u $-th road is written as
\begin{align}
R^\text{V}_{ud} \geqslant \log_2\left( 1+\dfrac{1}{\mathbb{E}\left[ \left( \gamma^\text{V}_{ud} \right)^{-1} \right]} \right)-\log_2 \myexp\sqrt{\dfrac{1}{\lambda}}Q^{-1}\left( \epsilon^\text{V}_{ud} \right).
\end{align}
where $ \lambda $ is equal to $ \tau_\text{SP}-\tau_\text{RP} $ in the RP scheme, while it is equal to $ \tau_\text{SP} = L^\text{V}B $ in the SP scheme. The proof of the RP is similar to that of the SP, thus the SINR expression of the RP is omitted here. Based on \eqref{VSPgamma}, we obtain
\begin{align}
\label{VSPgammarec}
& \left( \gamma^\text{V,SP}_{ud} \right)^{-1} = \sum_{i=1, i \neq d}^{D_u} \chi^\text{V,SP}_{ui} \dfrac{p^\text{V}_{ui}}{p^\text{V}_{ud}} \dfrac{q^\text{V}_{ui}}{q^\text{V}_{ud}} \dfrac{\left( \beta^\text{V2V}_{ud,ui} \right)^2}{\left( \beta^\text{V2V}_{ud,ud} \right)^2} + \sum_{j=1, j \neq u}^{4} \sum_{i=1}^{D_j} \chi^\text{V,SP}_{ji} \dfrac{p^\text{V}_{ji}}{p^\text{V}_{ud}} \dfrac{q^\text{V}_{ji}}{q^\text{V}_{ud}} \dfrac{\left( \beta^\text{V2V}_{ud,ji} \right)^2}{\left( \beta^\text{V2V}_{ud,ud} \right)^2} \notag \\
&\quad + \sum_{k=1}^{K} \chi^\text{V,SP}_{k} \dfrac{p^\text{C}_{k}}{p^\text{V}_{ud}} \dfrac{q^\text{C}_{k}}{q^\text{V}_{ud}} \dfrac{\left( \beta^\text{C2V}_{ud,k} \right)^2}{\left( \beta^\text{V2V}_{ud,ud} \right)^2} +\sum_{j=1}^{4} \sum_{i=1}^{D_j} \dfrac{\left( p^\text{V}_{ji} \right)^2}{\tau_\text{SP}p^\text{V}_{ud}q^\text{V}_{ud}} \dfrac{\left( \beta^\text{V2V}_{ud,ji} \right)^2}{\left( \beta^\text{V2V}_{ud,ud} \right)^2} + \sum_{k=1}^{K} \dfrac{\left( p^\text{C}_{k} \right)^2}{\tau_\text{SP}p^\text{V}_{ud}q^\text{V}_{ud}} \dfrac{\left( \beta^\text{C2V}_{ud,k} \right)^2}{\left( \beta^\text{V2V}_{ud,ud} \right)^2}.
\end{align}

First of all, let us calculate the mathematical expectation for \eqref{VSPgammarec} over $ \chi $, namely $ \mathbb{E}_{\chi}[( \gamma^\text{V,RP}_{ud} )^{-1}] $ and $ \mathbb{E}_{\chi}[( \gamma^\text{V,SP}_{ud} )^{-1}] $. According to the pilot allocation in \eqref{dist}, we have $ \mathbb{E}[\chi^\text{V,RP}_{ji}] = \mathbb{E}[\chi^\text{V,RP}_{k}] = 1/\tau_\text{RP} $ and $ \mathbb{E}[\chi^\text{V,SP}_{ji}] = \mathbb{E}[\chi^\text{V,SP}_{k}] = 1/\tau_\text{SP}, \forall k, (j,i) \neq (u,d) $~\cite{Verenzuela2018}, which leads to \eqref{VGamma}. 

Then, \eqref{VGamma} can be further averaged over the vehicle density and large-scale fading. For the $ d $-th V2V receiver on the $ u $-th road, the ideal case is to transmit signals with the maximum power in the interference-free scenario. Without loss of generality, the \textit{worst case} without resource allocation is thus considered here, namely the maximum interference scenario. In the RP scheme, let $ p^\text{V}_{ji} = q^\text{V}_{ji} = P^\text{V}_\text{max} $ and $ p^\text{C}_{k} = q^\text{C}_{k} = P^\text{C}_\text{max}, \forall k, (j,i) $, while in the SP scheme, let $ p^\text{V}_{ji} = \delta^\text{V}_{ji}P^\text{V}_\text{max}, q^\text{V}_{ji} = (1-\delta^\text{V}_{ji})P^\text{V}_\text{max} $ and $ p^\text{C}_{k} = \delta^\text{C}_{k}P^\text{C}_\text{max}, q^\text{C}_{k} = (1-\delta^\text{C}_{k})P^\text{C}_\text{max}, \forall k, (j,i), \forall \delta \in (0,1) $. Since $ f(x)=x(1-x), \forall x \in (0,1) $ is a concave function and $ f(0)=f(1)=0 $, $ pq=\delta(1-\delta)P^2_\text{max} $ gains the maximum value $ P^2_\text{max}/4 $ when $ \delta=1/2 $. Recall from $ \beta^\cdot_\cdot = \theta (d^\cdot_\cdot)^{-\alpha} $ that \eqref{VGamma} can be rewritten as
\begin{align}
\label{VSPGammarec}
\left( \Gamma^\text{V,SP}_{ud} \right)^{-1} &= \sum_{i=1, i \neq d}^{D_u} \dfrac{1}{\tau_\text{SP}} \dfrac{\left( d^\text{V2V}_{ud,ui} \right)^{-2\alpha}}{\left(d^\text{V2V}_{ud,ud} \right)^{-2\alpha}} + \sum_{j=1, j \neq u}^{4} \sum_{i=1}^{D_j} \dfrac{1}{\tau_\text{SP}} \dfrac{\left( d^\text{V2V}_{ud,ji} \right)^{-2\alpha}}{\left(d^\text{V2V}_{ud,ud} \right)^{-2\alpha}} \notag \\
&\quad +\sum_{j=1}^{4} \sum_{i=1}^{D_j} \dfrac{1}{\tau_\text{SP}} \dfrac{\left( d^\text{V2V}_{ud,ji} \right)^{-2\alpha}}{\left(d^\text{V2V}_{ud,ud} \right)^{-2\alpha}} + \sum_{k=1}^{K} \dfrac{2\psi^2}{\tau_\text{SP}} \dfrac{\left( d^\text{C2V}_{ud,k} \right)^{-2\alpha}}{\left(d^\text{V2V}_{ud,ud} \right)^{-2\alpha}},
\end{align}
where $ \psi = P^\text{C}_\text{max}/P^\text{V}_\text{max} $ is the ratio of the maximum power for V2V pairs and CUEs. Next, we calculate $ \mathbb{E}_{\{\rho, d\}}[( \Gamma^\text{V}_{ud} )^{-1}] $ for these two schemes. Based on Appendix~\ref{app:calc}, we can acquire the upper bounds of $ \mathbb{E}[d^{-2\alpha}] $ and $ \mathbb{E}[d^{2\alpha}] $. Furthermore, \eqref{LWRmodel} gives rise to $ \mathbb{E}[D(\rho)] = \sum_{d=0}^{+\infty} d \cdot \mathbb{P}[D(\rho)=d]=\frac{\rho A_\text{RL}A_\text{RW}}{2}, \forall u $.
With the fact that all V2V pairs and CUEs are i.i.d., we substitute the results of $ \mathbb{E}[d^{-2\alpha}] $, $ \mathbb{E}[d^{2\alpha}] $ and $ \mathbb{E}[D(\rho)] $ into \eqref{VSPGammarec} leads to \eqref{aveVGamma}.

\section{Calculations of Path Loss}
\label{app:calc}

\begin{figure}[!t]
	\centering
	\includegraphics[scale=0.8]{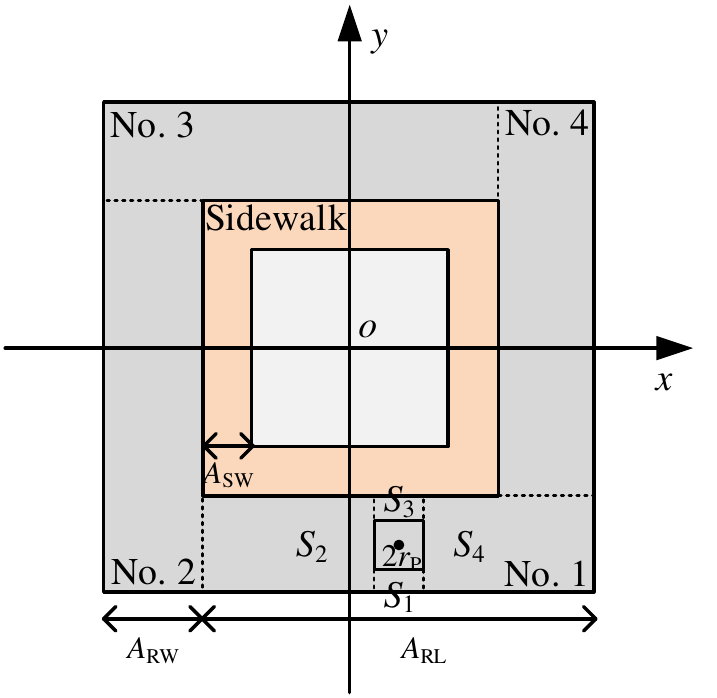}
	\caption{Cartesian coordinate for single-cell urban V2V URLLC system.}
	\label{fig_location}
\end{figure}

As shown in Fig.~\ref{fig_location}, a Cartesian coordinate is established for the single-cell urban V2V URLLC system. Without loss of generality, we number four roads in order to illustrate more clearly. Based on \eqref{LWRmodel}, we have the following distributions for the locations of all V2V pairs, i.e.,
\begin{align}
\label{Vdist}
\begin{cases}
x^\text{V}_{1d} \sim \text{unif}\left( -\dfrac{A_\text{RL}-A_\text{RW}}{2},\dfrac{A_\text{RL}+A_\text{RW}}{2} \right), \\
y^\text{V}_{1d} \sim \text{unif}\left( -\dfrac{A_\text{RL}+A_\text{RW}}{2},-\dfrac{A_\text{RL}-A_\text{RW}}{2} \right),
\end{cases} \text{for Road 1},
\end{align}
the distributions of the other roads are similar to that of Road 1, hence they are omitted. To ensure the convergence of $ \mathbb{E}[ ( d^\text{V2V}_{ud,ji} )^{-2\alpha} ] $, we need to set the protection region for V2V receivers. Since all V2V pairs are uniformly located at a plane, the protection region is set as a square around the V2V receiver with the length $ 2 r_\text{P} $, where $ r_\text{P} $ is related to the size of vehicle. According to \eqref{Vdist}, the probability density function (PDF) of $ d^\text{V}_{ud} $ is given by $ f_{d^\text{V}_{ud}}\left( x,y \right) = \frac{1}{\left( A_\text{RL}-2r_\text{P} \right)\left( A_\text{RW}-2r_\text{P} \right)}, \forall (u,d) $. Analogously, all CUEs are also uniformly located along the sidewalk, and the PDF of $ d^\text{C}_{k} $ is given by $ f_{d^\text{C}_{k}}\left( x,y \right) = \frac{1}{\left( A_\text{RL}-A_\text{RW}-A_\text{SW} \right)A_\text{SW}}, \forall k $.

The joint PDFs of $ d^\text{V2V}_{ud,ji} $ and $ d^\text{C2V}_{ud,k} $ can be written as $ f_{d^\text{V2V}_{ud,ji}}( x,y,z,w ) = f^2_{d^\text{V}_{ud}}, \forall (j,i), j \neq u $, $ f_{d^\text{C2V}_{ud,k}}( x,y,z,w ) = f_{d^\text{V}_{ud}}f_{d^\text{C}_{k}}, \forall k $, and $ f_{d^\text{V2V}_{ud,ui}}( x,y,z,w ) = f_{S_t}( z,w | x,y )f_{d^\text{V}_{ud}}, \forall i $, where $ f_{S_t}( z,w | x,y ), t \in \{1,2,3,4\} $ is the conditional PDFs of four subregions shown in Fig.~\ref{fig_location}. Notice that the above all PDFs are the upper bounds. Utilizing the upper bounds is convenient for calculations, and is also conducive to getting the lower bound of SE. In the following analysis, the abbreviations `N' and `P' denote the positive and negative polarities of exponents, respectively. Based on the symmetry of network topology, the average path loss for V2V pairs can be calculated as $ \mathbb{E}\left[ \left( d^\text{C2V}_{ud,k} \right)^{-2\alpha} \right] = \Omega^\text{C2V}_{\text{N}}, \forall k $, and
\begin{align}
\label{V2Vcalc}
\mathbb{E}\left[ \left( d^\text{V2V}_{ud,ji} \right)^{-2\alpha} \right] = \begin{cases}
\Omega^\text{V2V}_{\text{N},1}, &\qquad \text{for}~j = u, \forall i \neq d, \\
\Omega^\text{V2V}_{\text{N},2}, &\qquad \text{for}~j \perp u, \forall i, \\
\Omega^\text{V2V}_{\text{N},3}, &\qquad \text{for}~j \| u, \forall i,
\end{cases}
\end{align}
where $ j \perp u $ represents the fact that the $ j $-th road is connected (perpendicular) to the $ u $-th road, such as the relation between Rood 1 and Rood 2 in Fig.~\ref{fig_location}. $ j \| u $ represents the fact that the $ j $-th road parallels to the $ u $-th road, such as the relation between Rood 1 and Rood 3. Furthermore, due to the local characteristics of V2V communications, the distance between the transmitter and receiver of each V2V pair is set as $ r_\text{V} $~\cite{Ghazanfari2019}. Hence, we have $ \mathbb{E}\left[ \left( d^\text{V2V}_{ud,ud} \right)^{2\alpha} \right] = \iiiint\limits_{d^\text{V2V}_{ud,ud} \leqslant r_\text{V}} \left[ \left( x-z \right)^2+\left( y-w \right)^2 \right]^{\alpha} f_{d^\text{V2V}_{ud,ud}} \text{d}x\text{d}y\text{d}z\text{d}w = \Omega^\text{V2V}_{\text{P},1} $.

Similarly, the average path loss for CUEs is given by $ \mathbb{E}\left[ \left( d^\text{C2B}_{k} \right)^{-2\alpha} \right] = \Omega^\text{C2B}_{\text{N}} $, $ \mathbb{E}\left[ \left( d^\text{C2B}_{k} \right)^{2\alpha} \right] = \Omega^\text{C2B}_{\text{P}}, \forall k $, and $ \mathbb{E}\left[ \left( d^\text{V2B}_{ud} \right)^{-2\alpha} \right] = \iint \left( x^2+y^2 \right)^{-\alpha} \frac{1}{A_\text{RL}A_\text{RW}} \text{d}x\text{d}y = \Omega^\text{V2B}_{\text{N}}, \forall (u,d) $.

%
\bibliographystyle{IEEEtran}
\bibliography{IEEEabrv,Ref}
\end{document}